\documentclass[preprint,12pt]{elsarticle}
\usepackage{amsmath}       
\usepackage{amssymb}       
\usepackage{amsfonts}      
\usepackage{stackengine}   
\DeclareUnicodeCharacter{2212}{-} 

\usepackage{algorithm}
\usepackage{algpseudocode}

\usepackage{booktabs}      
\usepackage{tabularx}      
\usepackage{multirow}      
\usepackage{array}         
\usepackage{colortbl}      
\usepackage{hhline}        
\usepackage{diagbox}       
\usepackage{tabularray}    
\newcolumntype{P}[1]{>{\centering\arraybackslash}p{#1}} 

\usepackage{graphicx}      
\usepackage{wrapfig}       
\usepackage{subcaption}    
\usepackage{caption}       
\usepackage[font=footnotesize,labelfont=bf,textfont=sc]{caption} 
\usepackage{pgfplots}      
\pgfplotsset{compat=1.16}  

\usepackage[margin=3.0cm]{geometry} 
\usepackage{lipsum}         
\usepackage{comment}        
\usepackage{dirtytalk}      
\usepackage{rotating}       
\usepackage{wasysym}        
\usepackage{pifont}         

\DeclareMathAlphabet\mathzapf{T1}{pzc}{mb}{it} 

\usepackage{adjustbox}

\journal{Engineering Applications of Artificial Intelligence}
\date{}

\begin{document}
\begin{frontmatter}
\title{Recurrence Plot and Change Quantile-based Deep Supervised and Semi-supervised Protection for Transmission Lines Connected to Photovoltaic Plants}

\author[a]{Pallav Kumar Bera}
\author[b]{Samita Rani Pani}
\address[a]{Electrical Engineering, Western Kentucky University, Bowling Green, KY, USA}
\address[b]{School of Electrical Engineering, KIIT University, Bhubaneswar, Odisha, India}

\begin{abstract}
Conventional relays encounter difficulties in protecting transmission lines (TLs) connected to converter-based energy sources (CBESs) due to the influence of power electronics on fault characteristics. This article proposes a single-ended intelligent protection method for the TL segment between the grid and a Photovoltaic (PV) plant. The approach utilizes a Recurrence Matrix and an InceptionTime-based system to identify faults by using the mean change in quantiles of 3-phase currents. It determines the fault position and identifies the faulty phase. ReliefF feature selection is applied to extract the optimal quantile features. The scheme's performance is assessed under abnormal conditions, including faults and capacitor and load-switching events, simulated in Power Systems Computer Aided Design / Electromagnetic Transients Program (PSCAD/EMTDC) on the Western System Coordinating Council (WSCC) 9-bus system, with various fault and switching parameters. The scheme is also validated on the New England IEEE 39-bus system and in presence of partially rated converters. Additionally, the validation of the proposed strategy takes into account various conditions, including double-circuit line configuration, noise, series compensation, high-impedance faults, current transformer (CT) saturation, evolving and cross-country faults, remote and local faults, as well as variations in PV capacity, sampling frequency, and data window size. {To address label scarcity and improve generalization, semi-supervised learning paradigms including label spreading, label propagation, and self-training are integrated with the InceptionTime framework, enabling near-supervised performance with limited annotated fault data.} The results demonstrate that the approach is effective in handling different system configurations and conditions, ensuring the protection of TLs connected to large PV plants.
\end{abstract}


\end{frontmatter}

\section{INTRODUCTION}

In recent years, sustainable power generation (SPG) have gained substantial traction as an alternative to fossil fuels in electricity generation. This growth is expected to continue as governments, corporations, and individuals increasingly embrace SPG for their environmental and economic benefits. A considerable portion of SPG, including solar photovoltaic (PV) systems and wind farms (WFs), is integrated into the power grid by means of high-voltage transmission lines (TLs), using power electronic converters to facilitate energy transmission from remote locations.

The introduction of SPG has fundamentally altered traditional power system topologies, leading to bidirectional power flows and varied fault current levels. To maintain grid stability, modern grid codes impose fault ride-through (FRT) requirements on these converter-based energy sources (CBESs), which influence fault current characteristics due to the intermittent nature of the resources and their reliance on power electronic interfaces \cite{singh2018}. Unlike traditional synchronous generators, the fault current behavior in CBES is governed by specific  control strategies, inverter parameters, and the overall fault conditions in the system, among other factors.

The increasing integration of CBESs, such as large-scale PV and wind installations, presents substantial challenges to existing protection schemes, including distance, directional, differential, and overcurrent protection. These conventional schemes struggle to maintain dependability and security due to the new fault dynamics from CBESs, including lower fault currents to reduce thermal stress and lower negative-sequence currents to prevent bus capacitor overvoltage.

Studies have highlighted the impact of PV integration on traditional distance protection used in TLs.
The low and unique fault current characteristics from PV can lead to misoperations in distance relays \cite{liang}. 
Oscillations in apparent impedance caused by CBES-injected currents can result in overreach and underreach issues for phase distance elements \cite{ritwik}. The impedance behavior of CBESs under different FRT requirements is analyzed highlighting differences between measured and actual fault impedances due to control strategies in \cite{hosiyar2}. The presence of CBESs also impacts negative-sequence directional elements, which are crucial for phase and ground distance protection \cite{kou2020}. The misoperation of negative-sequence
overcurrent  and  directional negative-sequence overcurrent relays were studied in \cite{haddadi2021}.
In differential protection schemes, the integration of large-scale PV plants introduces reliability concerns due to disparities in short-circuit characteristics between PV inverters and traditional synchronous machines \cite{pvqiang}. 
Sensitivity challenges in differential protection of TLs with CBESs have been documented in \cite{yang}. Overcurrent relay performance is impacted by distortions in fault signals and shifts in angular relationships between voltages and currents, leading to challenges in fault direction determination, localization, and prevention of false tripping \cite{kumar}. Further, high-impedance fault detection becomes increasingly difficult in systems with integrated PV, as highlighted in \cite{kavi}. 

To address these challenges, alternative protection methods have been proposed. An enhanced distance protection scheme based on zero-sequence impedance and time delay was suggested in \cite{fang2019} to address impedance measurement distortions in CBES-integrated systems. A current differential protection approach was developed in \cite{saber22} using phase current signs at both ends of the TL connected to WFs. Transient current correlation-based differential protection \cite{JIA2018} and  positive sequence-based distance protection independent of plant parameters \cite{GHORBANI2023} are explored for CBES-integrated systems.  Fault detection, classification and location are performed using the magnitude ratios of the line end phase currents and apparent power in \cite{singh}. These methods aim to enhance fault detection sensitivity and reliability in TLs connected to CBES.

To further support protection reliability in complex scenarios, recent efforts have turned to machine learning (ML)-based protection schemes, which adapt dynamically to system conditions. A hybrid  fuzzy and  random forest (RF) using combined linear trends (CLT) was used in \cite{pallavpv} to detect and classify faults in grid-connected PVs. However, its application is limited by high sampling frequency requirements.  Techniques involving positive-sequence current and empirical mode decomposition (EMD) with  RF classifiers have been applied for fault detection in lines with TCSC compensation \cite{sauviktcsc21}.  {However, it primarily focuses on DFIG-based wind farms and may not readily extend to PV-integrated systems or hybrid renewable sources. Support Vector Machines (SVM) have been applied to estimate impedance in PV-integrated TLs under variable infeed conditions \cite{pvsvm}. Fault detection using discrete wavelet transform combined with SVM-based classification in renewable-energy-penetrated microgrids has been reported in \cite{BHAUMIK2025}. Similarly, ensemble tree-based approaches for fault detection in inverter-based generator systems have been presented in \cite{ADAPTIVE_TREE_2025,TREE_2022}. A deep learning–based protection algorithm employing two-dimensional spatial current trajectory imaging for PV-connected transmission lines has also been proposed in \cite{Dl_pv_2025}. However, these studies explore only a limited set of features and do not address validation under many critical scenarios, thereby restricting the robustness and generalizability of the methods.}

Due to the unique fault current characteristics that challenge existing protection techniques, it is essential to develop new protection schemes or enhance existing ones with a more comprehensive exploration of feature sets.
 {This article proposes a novel integration of recurrence matrix representations of change quantiles (RMCQs) with an InceptionTime-based classifier to achieve intelligent protection of transmission lines connected to PV plants.}
The primary contributions of this study are as follows:

\begin{itemize} \item A unique RMCQ-based intelligent technique is proposed for fault detection, fault localization, and identification of faulty phases. The method is tested across various scenarios, accounting for parameters influencing fault current behaviors.
\item The RMCQ-based protection approach is validated on the IEEE 39-Bus New England Test System and with partially rated converters. \item The proposed scheme is tested under challenging conditions, including noise, local and remote faults, high impedance faults, double circuit configurations, CT saturation, series compensation, evolving faults, cross-country faults, and with variations in PV capacity, sampling frequencies, and window sizes.
\item {A novel integration of graph-based (label spreading, label propagation) and confidence-based (self-training) semi-supervised learning techniques with the InceptionTime classifier is introduced, enabling reliable fault detection with minimal annotated data.}
\item The dataset containing fault and other transient data is made available on a public research data platform  {\cite{datapv}}.

\end{itemize}

The structure of the article is as follows: Section II presents the modeling and simulation of faults, load-switching, and capacitor-switching events on the WSCC 9-bus system. Section III details the RMCQ-based intelligent protection scheme, including the change quantile (CQ) features, ReliefF algorithm, time series imaging techniques, and the InceptionTime classifier. Section IV discusses the results of fault detection, location, and phase selection. In Section V, the impacts of noise, sampling frequency, series compensation, window size, CT saturation, PV unit variations, double circuit lines, evolving and cross-country faults, high impedance faults, and local and remote-end faults are analyzed. {Semi-supervised learning approaches are employed to improve data efficiency under limited annotation scenarios in Section VI.} Validation of the proposed method on the IEEE 39-bus system and with partially rated converters is performed in Sections VII and VIII, respectively. A comparative analysis with recent studies is provided in Section IX. Finally, Section X summarizes the conclusions. 

\begin{figure}[ht]
\centering
\captionsetup{justification=centering,textfont=small}
\includegraphics[width=4.4 in, height=2.3 in]{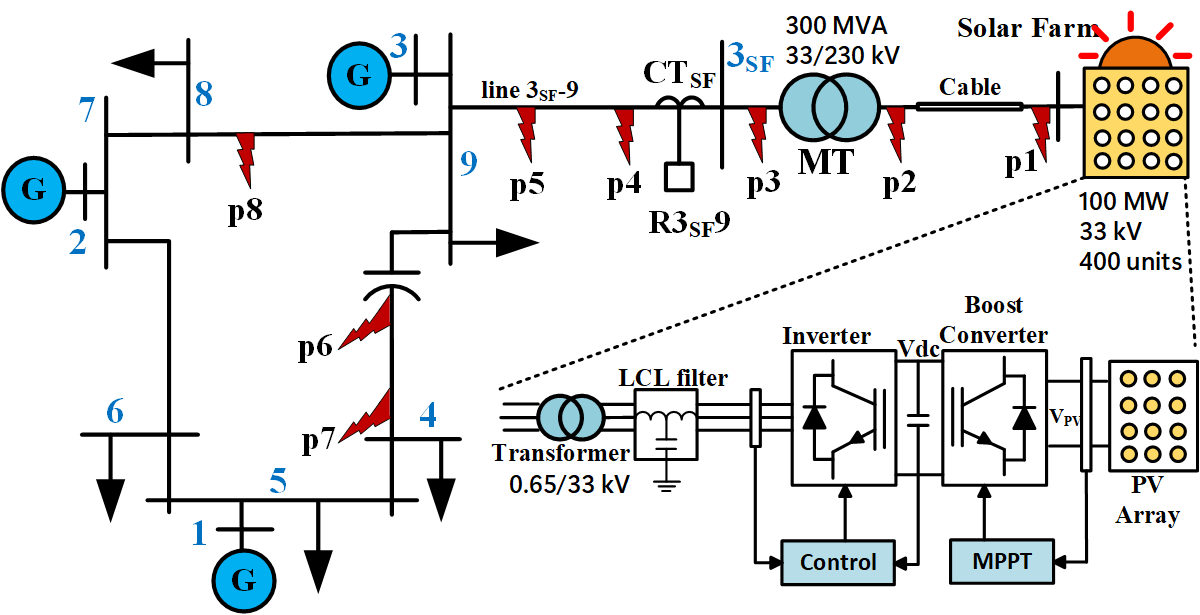}
\vspace{-0.1cm}
\caption{WSCC 9-Bus System with PV Plant at bus-9.}
\label{9busPV}
\vspace{-1mm}\end{figure}

\section{Modeling and Simulation for PV}

A 100 MW photovoltaic (PV) system is integrated into the WSCC 9-bus system at bus-9, as shown in Fig.\ref{9busPV}. The TL connecting the PV plant and the grid spans 100 km, with a positive-sequence impedance of (1.0 + 30.0j) $\Omega$ and a zero-sequence impedance of (33.0 + 110.0j) $\Omega$. To enhance power transfer capability and operational efficiency, TL 4-9 is series compensated.
Faults are introduced at eight distinct positions by adjusting variables: priority mode, fault impedance, fault type, and fault onset moment. The generator at bus-3 is alternately switched on and off to assess the impact of infeed, particularly in scenarios involving capacitor and load addition. Tables \ref{parameters1} and \ref{parameters2} summarize the parameters and values utilized for simulating these fault scenarios, along with load and capacitor switching cases.

\begin{table}
\renewcommand{\arraystretch}{1.15}
\setlength{\tabcolsep}{2 pt}
\centering
\caption{{Parameters and values for fault simulations} }\label{parameters1}
\footnotesize
\begin{tabular}{|l|l|} 
\hline
\multicolumn{2}{|c|}{{\cellcolor[rgb]{0.89,0.89,0.89}}\textbf{Fault Events}}                                                            \\ 
\hline
Priority Modes         & Active Power (P) and Reactive Power (Q) (2)                                                                                            
                                          \\ 
                                          \hline
Fault positions           & p1, p2, p3, p4, p5, p6, p7, p8     (8)                                                                                \\ 
\hline
Fault types               & $ag, bg, cg, ab, bc, ca, abg, bcg, cag, abcg$ (10)                                                            \\ 
\hline
Fault impedances($R_f$) & 0.01, 1, 10$\Omega$ (3)                                                                                      \\ 
\hline
Fault onset moments & 9.0, 9.00334, 9.00668, 9.01002, 9.01336, 9.0167s (6)                                 \\ 

\hline
\multicolumn{2}{|l|}{{\cellcolor[rgb]{0.949,0.949,0.949}}Total fault cases =$ 2\times 8\times 10 \times 3 \times 6$ = 2880}    \\ 
\hline
\end{tabular}
\end{table}

\begin{table}[htpb]
\renewcommand{\arraystretch}{1.15}
\setlength{\tabcolsep}{6 pt}

\caption{{Parameters and values for other transient simulations} }\label{parameters2}
\footnotesize
\centering
\begin{tabular}{|l|l|} 
\hline
\multicolumn{2}{|c|}{{\cellcolor[rgb]{0.89,0.89,0.89}}\textbf{Non-Fault Events }}                                                        \\

\hline
Priority Modes         & Active Power (P) and Reactive Power (Q) (2)                                                                                             \\ 
\hline
Generator (at bus-3)     & Engaged / Disengaged (2)                                                                                   \\ 

\hline
Positions                 & Bus-4, 8, 9    (3)    \\
\hline
Capacitor/Load capacity    & 4 different ratings                                                                                                            \\ 
\hline
Switching moments          & 9.0, 9.00069, 9.00138, $\ldots$, 9.01656s (25) \\
\hline
\multicolumn{2}{|l|}{Capacitor-energization cases = $2\times 2\times 3 \times 4 \times 25$ = 1200}                                          \\ 
\hline
\multicolumn{2}{|l|}{Load-addition cases = $2\times 2\times 3 \times 4 \times 25$ = 1200}                                               \\ 
\hline
\multicolumn{2}{|l|}{{\cellcolor[rgb]{0.949,0.949,0.949}}Total non-fault cases = 2400}                                                   \\
\hline
\end{tabular}
\end{table}

\begin{figure}[ht]
\captionsetup{justification=centering,textfont=small}
\centerline{\includegraphics[width=4.0 in, height= 1.60 in]{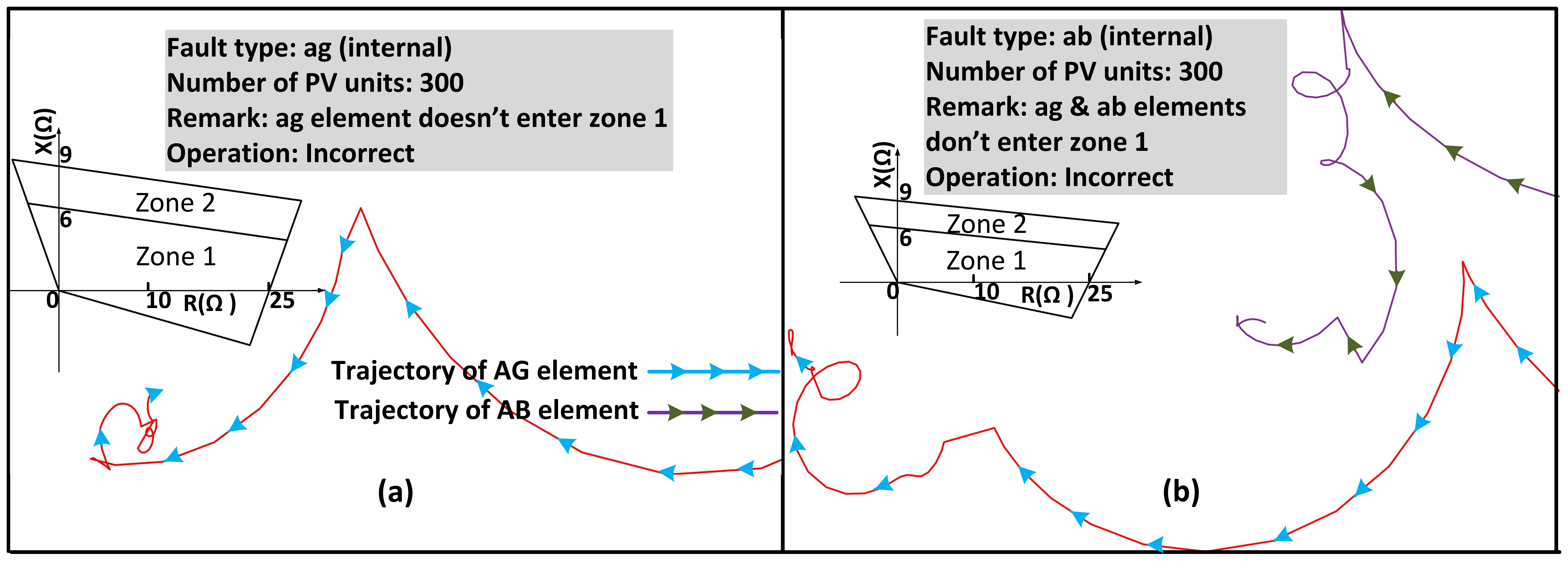}}
\caption{Impedance calculated by AG \& AB elements of R$3_{SF}$9}
\label{relayop}
\end{figure}

\textit{Maloperation of distance relays:} The maloperation of impedance based distance relay R$3_{SF}$9 is illustrated by the fault behavior in the impedance plane. For an $ag$ fault at position p5  within zone 1, the relay fails to operate in case 1 (PV units = 300, fault impedance $R_f$=10$\Omega$, and priority mode P). Similarly, it remains inactive during an $ab$ fault at the same location in case 2 (PV units = 300, $R_f$=10$\Omega$, priority mode P)  {(Fig. \ref{relayop})}. The procedure for determining the impedance trajectory of the quadrilateral distance relay adopted in \cite{hosiyar1} is outlined in Algorithm 1.

\begin{center}
\begin{adjustbox}{width=0.8\textwidth}
\begin{minipage}{\textwidth}
\begin{algorithm}[H] 
\small
\caption{Distance Relay Impedance Trajectory}
\begin{algorithmic}[1]
\State \textbf{Input:} Voltages $v_a$, $v_b$, $v_c$ \& currents $i_a$, $i_b$, $i_c$ sampled at 7680 Hz.
\State \textbf{Filter Signals:} Apply 5th-order Butterworth filter with cutoff of 400 Hz.

\[
\text{[filter\_b, filter\_a]} = \text{butter}(5, 400 / (7680 / 2))
\]

\State \textbf{Downsample Data:} Reduce sampling rate by factor of 4 to 1920 Hz.
\State \textbf{Compute Fundamental Components:} Apply FFT to 32-sample windows and extract the 60 Hz components:

\[
v_a{'} = 2 \cdot \text{FFT}_{va}(2) / 32, \quad \text{similarly for } v_b{'}, v_c{'}, i_a{'}, i_b{'}, i_c{'}.
\]

\State \textbf{Calculate Apparent Impedances:} Compute zero-sequence current and phase impedances:

\[
\hspace{-5mm}I_0\!=\!\frac{i_a{'} + i_b{'} + i_c{'}}{3},
Z_a\!=\!\frac{v_a{'}}{i_a{'}\! +\! 3K_0I_0},Z_b\!=\!\frac{v_b{'}}{i_b{'} \!+\! 3K_0I_0},Z_c\!=\!\frac{v_c{'}}{i_c{'}\! + \!3K_0I_0}
\]
\[
\hspace{-5mm} K_0\! =\! \frac{Z_0 - Z_1}{3Z_1},with\ Z_1\! =\! 0.23 + j7.6 \, \Omega, Z_0\! =\! 8.19 + j27.55 \, \Omega \].

\State \textbf{Define Quadrilateral Relay Zones:}
For line impedance of $Z_{\text{line}} = 1.0 + j30.0 \Omega$ and line length 100 km:

\[
Z_{\text{Zone 1}} = 0.8 \cdot Z_{\text{line}}, \quad Z_{\text{Zone 2}} = 1.2 \cdot Z_{\text{line}}
\]

\State Construct relay boundaries based on resistance and reactance thresholds.

\State \textbf{Impedance Trajectories:} Plot $(R, X)$ for $Z_a$, $Z_b$, $Z_c$, and line-to-line impedances:

\[
Z_{ab} = \frac{v_a{'} - v_b{'}}{i_a{'} - i_b{'}}, \quad \text{similarly for } Z_{bc}, Z_{ca}.
\]

\end{algorithmic}
\end{algorithm}
\end{minipage}
\end{adjustbox}
\end{center}

\vspace{-1mm}
\section{Suggested Protection}
\subsection{Outline of Protection Framework}
\begin{figure}[ht]\vspace{-1mm}
\captionsetup{justification=centering,textfont=small}
\centerline{\includegraphics[width=3.65 in, height= 3.05 in]{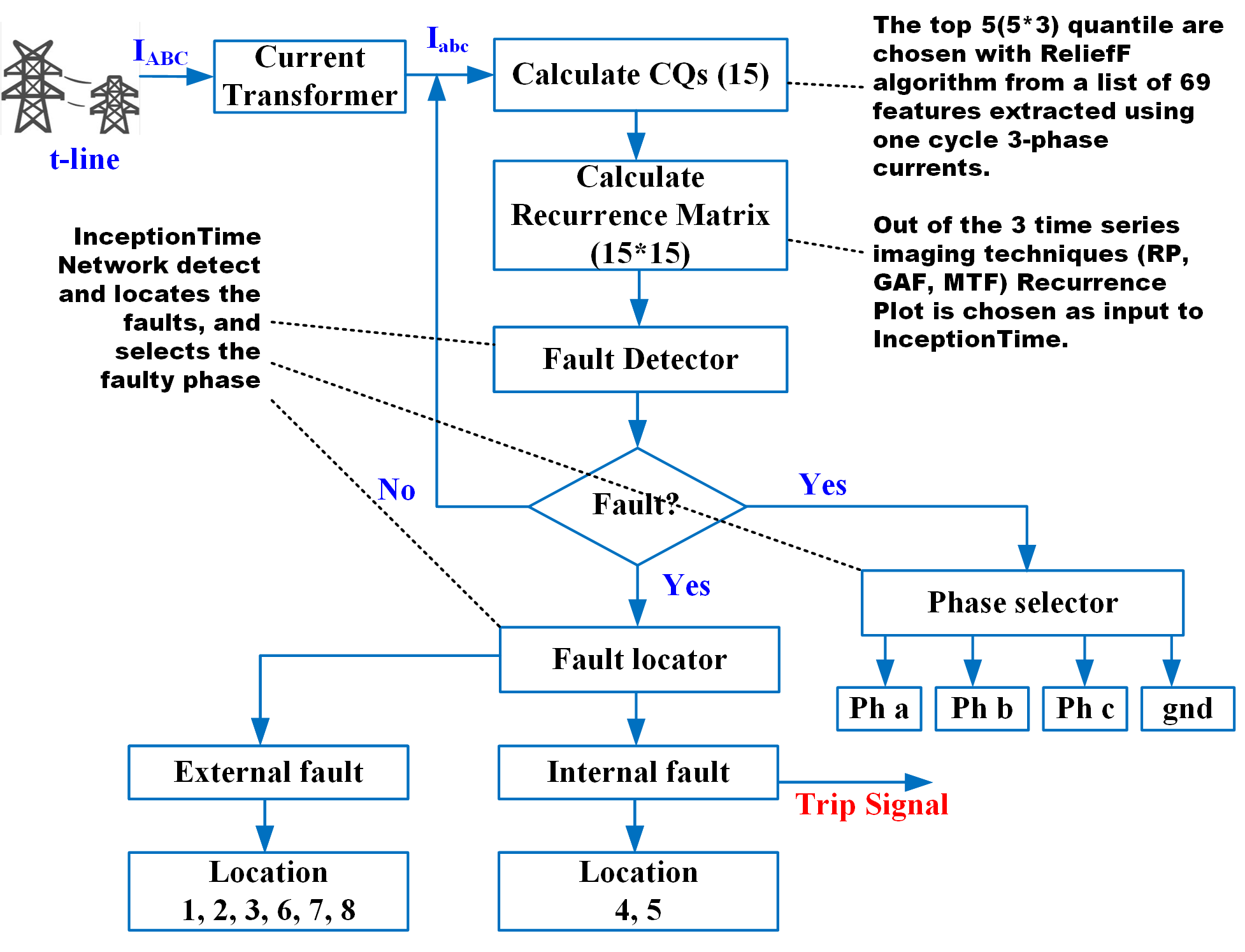}}
\caption{Flowchart of the suggested protection scheme.}
\label{flowdiagram}
\end{figure}

The suggested protection mechanism is illustrated in Fig. \ref{flowdiagram}. First, the 3-phase currents are recorded by the $CT_{SF}$ at the PV side of the TL under consideration (line $3_{SF}$-9). The recorded 1-cycle 3-phase currents are used to extract the Recurrence Matrix of Change Quantiles (RMCQ).  
Second, the RMCQ-trained InceptionTime Network is used for fault detection. Third, line $3_{SF}$-9 is tripped if the fault locator identifies the recorded transient currents as an internal fault (position p4 \& p5). Fourth, the faulty phase is determined. The first stage includes data pre-processing and feature extraction. ReliefF feature selection is used to rank the CQ features which are then used to form the recurrence matrix.

\vspace{-1mm}
\subsection{Change Quantiles}
The currents in a 3-phase relay system can be analyzed through various statistical features, which provide valuable insights into the dynamics of fault currents.
Quantiles are a versatile tool in statistics, used across various fields to analyze and interpret temporal patterns. However, their application in detecting power system events remains limited. Quantiles were employed to differentiate between faults and non-faults in interconnected systems, demonstrating promising results in \cite{systempallav}.
Change quantile (CQ) computes the aggregate of absolute values of consecutive changes in the time series within the range defined by the constant values $h$ and $l$.  {These features capture the current mutation rate induced by faults, reflecting the abrupt transients caused by inverter current-limiting actions and changes in system impedance.} The aggregates can be mean ($\mu$), median ($M$), standard deviation ($\sigma$), or variance ($\sigma^2$). In this way, 69 quantile features are extracted (Table \ref{featuretable}).
Mean CQ ($\mu CQ$) is defined using equation (\ref{mcq}).
\vspace{-1mm}

\begingroup
\small
\begin{equation}
 {\mu CQ =\frac{1}{n-1}\cdot{\sum_{t=1}^{n-1} |I_{ph}(t+1) - I_{ph}(t)| }\label{mcq}}
\end{equation}

\endgroup
\vspace{-1mm}
 {where, $n$  represents the number of sample points in the time series between $h$ and $l$, and $I_{ph}(t)$ refers to the phase currents at time step $t$.}

\textit{Sample Calculation of $\mu CQ$}: Using equation~(\ref{mcq}), the $\mu CQ$ of a time series $t$ = [-0.4,\ -0.2,\ -0.1,\ 0.5,\ 0.0] is calculated as follows:

\begin{align*}
\mu CQ   &= \frac{|-0.2 + 0.4| + |-0.1 + 0.2| + |0.5 + 0.1| + |0.0 - 0.5|}{4} \\
   &= \frac{0.2 + 0.1 + 0.6 + 0.5}{4} \\
   &= 0.35 \quad \text{for } h = 1 \text{ and } l = 0
\end{align*}

\begin{table}[ht]
\footnotesize
\renewcommand{\arraystretch}{1.2}
\setlength{\tabcolsep}{1 pt}
\centering
\captionsetup{justification=centering}
\caption{{Summary of quantile-based features extracted} }\label{featurelist}
\vspace{-1mm}
\label{featuretable}
\begin{tabular}{|c|c|c|}
\hline
\rowcolor[rgb]{0.91,0.91,0.91}$Features$       & $Description$                                       &     $Parameters$                                                                              \\ \hline

CQ (1-60) & \begin{tabular}[c]{@{}c@{}} $\mu$, $\sigma^2$,
$\sigma$, and  $M$ of\\$  {|I_{ph}(t+1) - I_{ph}(t)|} \forall t\!\in\! \mathbb{N}| t\leq n-1$ \\ betweeen $h$ and $l$  \end{tabular}                                             & \begin{tabular}[c]{@{}c@{}}$h$=1, 0.8, 0.6, 0.4, 0.2\\ $l$=0.8, 0.6, 0.4, 0.2, 0\\ Total=(5+4+3+2+1)$\times$4=60 \end{tabular}                             \\ \hline
Quantile  (61-69)      & \begin{tabular}[c]{@{}c@{}}Value of phase current 
\\$I_{ph}$ $\geq$ p\% of ordered $I_{ph}$ values\end{tabular}  & \begin{tabular}[c]{@{}c@{}}p = 0.1, 0.2, 0.3, 0.4,\\ 0.5, 0.6, 0.7, 0.8, 0.9\\Total=9\end{tabular} \\ \hline
\end{tabular}

\end{table}

\subsection{Relief Feature Selection}
\begin{figure}[ht]\vspace{-1mm}
\captionsetup{justification=centering,textfont=normal}
\centerline{\includegraphics[width=4.25 in, height= 2.13 in]{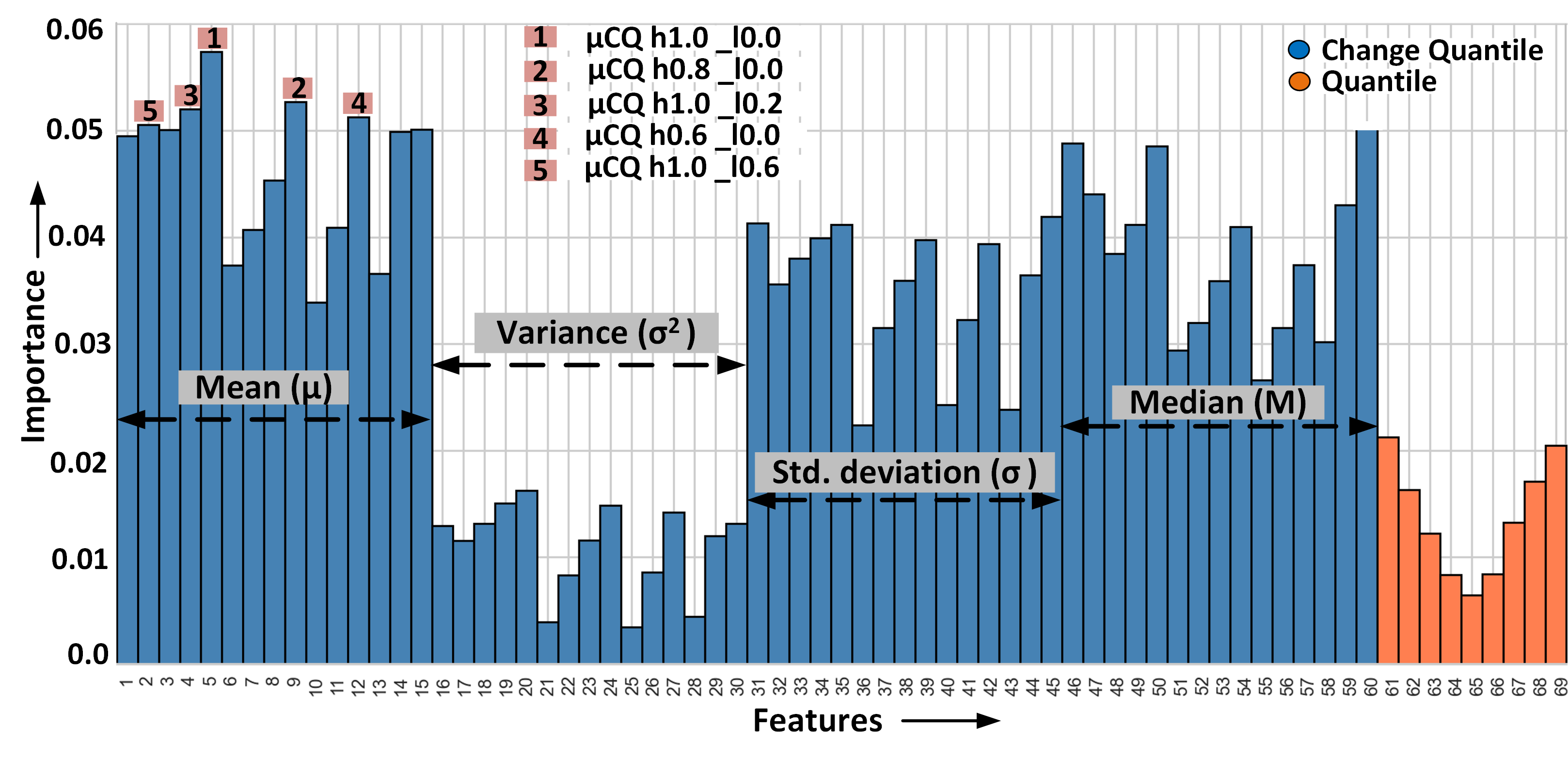}}
\caption{Feature score and ranking using ReliefF}
\label{relieff}
\vspace{-1mm}\end{figure}

\textit{Relief} is a feature selection technique that assesses the significance of features by measuring their ability to differentiate between instances belonging to different classes \cite{kira1992feature}. The Relief algorithm steps are outlined in Algorithm 2:

\begin{center}
\begin{adjustbox}{width=0.8\textwidth}
\begin{minipage}{\textwidth}
\begin{algorithm}[H] 
\small
  \caption{Relief Algorithm}
    \begin{algorithmic}[1]
        \State Initialize weights: $W_j$$\gets$0 for each feature j = 1, 2,$\ldots$, M
        \For{each iteration \( i = 1, 2, \ldots, T \)} 
            \State Randomly select an instance \( x_i \)
            \State Find the nearest hit \( \text{NH}(x_i) \)
            \State Find the nearest miss \( \text{NM}(x_i) \)
            \For{each feature \( j \)}
            \State Update weights 
            \[   W_j \gets W_j - \frac{(x_i^j - \text{NH}(x_i)^j)^2}{N} + \frac{(x_i^j - \text{NM}(x_i)^j)^2}{N}
                \]
            \EndFor
        \EndFor
        \State Normalize weights if necessary
\end{algorithmic}
\end{algorithm}
\end{minipage}
\end{adjustbox}
\end{center}
The algorithm selects the top 5 most important features (15 across 3 phases) from a set of 69 quantile-based features listed in Table \ref{featuretable}. The bar plot in Fig.\ref{relieff} illustrates the feature importance and ranking results from ReliefF. It is evident that the CQ features (features 1-60) exhibit higher importance than the quantile features (features 61-69), indicating a stronger contribution to the model's performance. Notably, the top 5 selected features use the mean ($\mu$) as the aggregate. Fig.\ref{relieff} also highlights these top 5 CQ features.

\subsection{Time Series Imaging} The transient time-series data is transformed into visual representations using Gramian Angular Field, Markov Transition Field, and Recurrence Plot techniques to capture and analyze the temporal patterns.

\subsubsection{Gramian Angular Field (GAF)} The GAF technique encodes time-series data into a two-dimensional matrix \cite{gaf}. The process begins by normalizing the time-series data to reduce the impact of outliers and extreme values. Then, the normalized time series is mapped from cartesian to polar coordinates to maintain temporal dependencies. Here, variations in the data amplitude are represented as angular changes in the polar coordinate system. This approach establishes temporal relationships between pairs of points by calculating the cosine summation of their moments, resulting in the Gramian Angular Summation Field (GASF), as illustrated in equation (2).

\begingroup
\small
 {\begin{equation}
GASF = \begin{bmatrix}
\cos(\alpha_1 + \alpha_1) & \cdots & \cos(\alpha_1 + \alpha_N) \\
\cos(\alpha_2 + \alpha_1) & \cdots & \cos(\alpha_2 + \alpha_N) \\
\vdots & \ddots & \vdots \\
\cos(\alpha_N + \alpha_1) & \cdots & \cos(\alpha_N + \alpha_N)
\end{bmatrix}
\end{equation}}
\endgroup


\subsubsection{Markov Transition Field (MTF)}
MTF is a technique used to encode time series data into a 2D matrix based on the probabilistic transitions between different quantile regions of the time series \cite{gaf}. First, the time series is discretized into predefined levels, typically using quantiles to ensure equal distribution of data across states. Then, a Markov transition matrix is constructed, representing the probability of transitioning from one state to another in consecutive time steps.

To capture temporal information, the Markov transition probabilities are projected onto the original time series as a field, creating the MTF. The matrix entry at position (i, j) in the MTF indicates the transition probability between the states corresponding to time points i and j. This field captures both local and global time dependencies within the sequence, making it suitable for time series analysis using image-based models.

\subsubsection{Recurrence Plot (RP)}

RPs, originally introduced by Eckmann \cite{rp}, are a popular method for showing how states reoccur within a time series. Conventionally, they use a binary matrix to depict recurring states of a dynamical system. A more thorough comprehension of the time series structure is provided in this study, though, as the recurrence matrix is built using actual Euclidean distances between states.
The recurrence plot for a given time series \(\{t_p\}_{p=1}^N\) is made by comparing pairs of points from the phase space trajectory. The time series is embedded using delay embedding into a higher-dimensional space to generate the phase space trajectory.
The Euclidean distance between states \(t_p\) and \(t_q\) is represented by element \(M_{p,q}\) in the recurrence matrix \(\mathbf{M}\) of this modified recurrence plot, which is a \(N \times N\) matrix:

\begingroup
\small
\begin{equation}
 {M_{p,q} = \| \mathbf{t}_p - \mathbf{t}_q \|, 1 \leq p,q \leq N}
\end{equation}
\endgroup

\noindent where the embedded vector is represented by \(\mathbf{t}_p = (t_p, t_{p+\delta}, \dots, t_{p+(r-1)\delta})\), the time delay is represented by \(\delta\), the embedding dimension is represented by \(r\), and the Euclidean norm is represented by \(\| \cdot \|\). To determine the recurrence matrix from a time series \(\{t_p\}_{p=1}^N\), the embedding dimension \(r\) and time delay \(\delta\) are selected, the vectors \(\mathbf{t}_p\) are constructed for \(p = 1, 2, \dots, N - (r-1)\delta\), and the pairwise Euclidean distances \(\|\mathbf{t}_p - \mathbf{t}_q\|\) are calculated. Euclidean distances between each pair of points in a time series \(\mathbf{t} = [0.15, 0.08, -0.01]\) are calculated as follows:

1. distance $s_{12}$ is:
$|t_1$ - $t_2|$ = $|0.15$ - $0.08|$ = 0.07

2. distance $s_{13}$  is:
$|t_1$ - $t_3|$ = $|0.15$ - (-$0.01)|$ = 0.16

3. distance $s_{23}$  is:
$|t_2$ - $t_3|$ = $|0.08$ - (-$0.01)|$ = 0.09

The distance matrix \(\mathbf{M}\) that is produced can be written as follows:

\begingroup
\small
\begin{equation}
\mathbf{M} =
\begin{bmatrix}
s_{11} & s_{12} & s_{13} \\
s_{21} & s_{22} & s_{23} \\
s_{31} & s_{32} & s_{33}
\end{bmatrix}
=
\begin{bmatrix}
0 & 0.07 & 0.16 \\
0.07 & 0 & 0.09 \\
0.16 & 0.09 & 0
\end{bmatrix}
\end{equation}
\endgroup

\begin{figure}[ht]\vspace{-1mm}
\captionsetup{justification=centering,textfont=small}
\centerline{\includegraphics[width=4.2 in, height= 2.55 in]{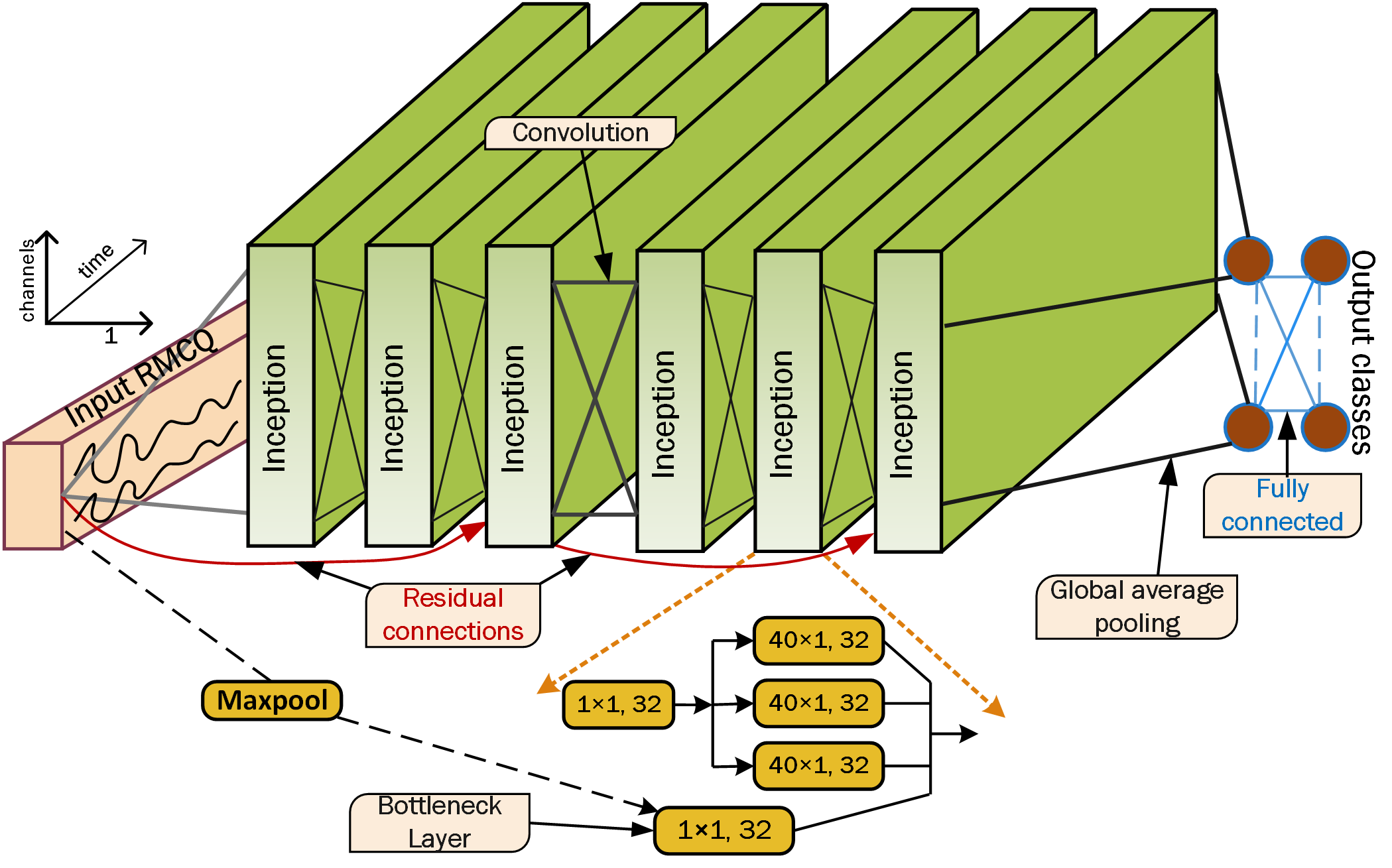}}
\caption{Architecture of InceptionTime Deep Learning Network}
\label{it}
\vspace{-1mm}\end{figure}

\vspace{-1mm}
\subsection{ InceptionTime (ICT) Network}

 {Recent works in intelligent systems have highlighted the potential of attention-based architectures, such as \cite{mixstyle} for machinery fault diagnosis, noise-robust transformer variants \cite{waveformer} for temporal sequence modeling under noisy conditions, and point spatio-temporal transformers \cite{psttransformer} for dynamic 3D point cloud video understanding. In contrast, this study employs recurrence–quantile imaging combined with an ICT classifier, offering a computationally efficient alternative for power system protection, where both robustness and speed are critical.}


The ICT classifier builds upon the success of inception-based networks in various computer vision applications. In \cite{autopallav}, the ICT framework was successfully applied to detect, localize, and classify faults for a transmission line connected to CBES. It is a highly accurate, scalable, and fast deep learning ensemble for time series classification. It is characterized by 2 residual blocks, each containing 3 inception modules. An inception module uses different filter lengths on time series inputs in parallel to extract relevant patterns at different temporal scales.
Each residual block's input is sent to the following block via a linear link in order to alleviate the gradient vanishing issue, enabling stable training of deep networks. A global average pooling layer follows the residual blocks, which computes the average of the multivariate time series output. A completely connected layer with a softmax activation function completes the final classification. 
The ICT architecture as described  in ref. \cite{Ismail_Fawaz_2020} with minute changes in filter length is used in this work (Fig. \ref{it}). Each of its five distinct, randomly initiated inception networks is created by cascading six inception modules and applying the receptive field idea to effectively capture patterns in temporal data. 
3 sets of filters with 32 filters each, and a kernel size of 40 are used in the convolutional layers in the inception modules. 
A bottleneck layer with size 32 is used to reduce the dimensionality of the input before applying convolutions, making the model more efficient in terms of computation and memory usage.
Maxpooling is also applied in parallel to the convolutional layers within each module to further reduce the input's dimensionality. 
This structure ensures that the model can learn diverse features from the time series data.
The model employs  Adam optimizer to accelerate convergence, Glorot uniform initializer to randomly initialize the weights, and  categorical cross-entropy loss function in order to classify many classes. It is trained for 1500 epochs with a mini batch size of 64 to improve generalization.

 {The proposed scheme employs multiple algorithms in a complementary sequence, where each step serves a distinct purpose. CQs are used to capture subtle variations in inverter-based fault currents, while ReliefF ensures the selection of the most discriminative and compact feature set. Time-series imaging (RP, GAF, MTF) transforms these features into structured 2D representations suitable for deep learning; among them, RP proved most effective. {When projected into the recurrence domain, the CQ sequence exhibits clustered high-distance regions signifying fault-onset dynamics—offering a physically interpretable signature of energy exchange during the transient.} Finally, the ICT network is adopted for its proven ability to extract multi-scale temporal patterns from image-based data with high accuracy and robustness. Together, this pipeline ensures dependable, generalizable, and computationally efficient.}

\section{Results of Fault Detection, Fault localization, and Phase Selection on IEEE 9-bus system}
The results of \textit{supervised learning} with ICT for fault detection (FD), phase selection (PS), and fault location (FL) phases on the WSCC 9-bus system's  are examined in this section.
To prevent overfitting and ensure that models can generalize to new, unseen data, the fault and transient dataset is divided into two subsets: a training set and a test set, using a 7:3 split ratio.
 Accuracy ($\mathcal{A}$) is used as the typical metric to measure the performance of the classifiers where $\mathcal{A} = \frac{\mathcal{N_{\text{correct}}}}{\mathcal{N_{\text{Total}}}}$ with $\mathcal{N_{\text{correct}}}$ being number of correctly predicted labels (either positive or negative) and $\mathcal{N_{\text{Total}}}$
being the total number of data points.
Furthermore, SMOTE (Synthetic Minority Over-sampling Technique) \cite{smote} is used to create synthetic samples of the minority class in order to mitigate the problem of unbalanced datasets.

\begin{figure}[ht]\vspace{-1mm}
\captionsetup{textfont=small}
\centerline{\includegraphics[width=3.0 in, height= 1.5 in]{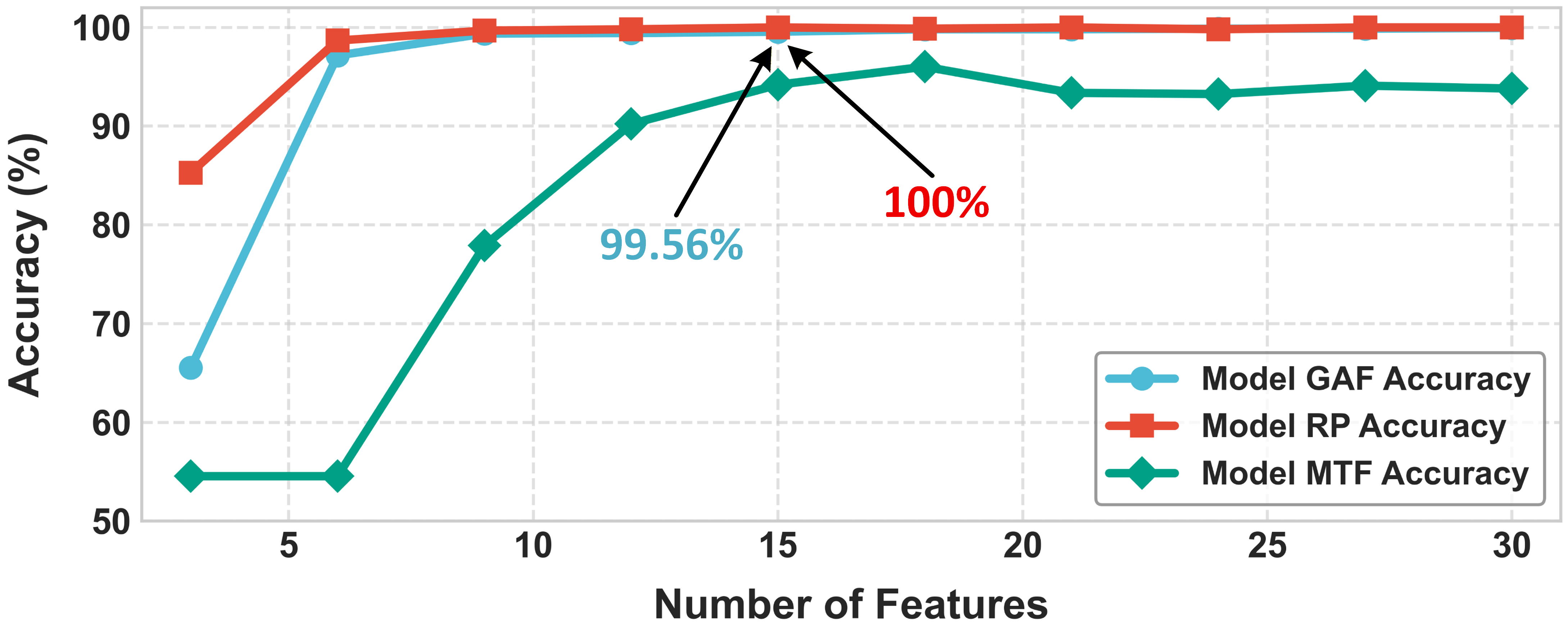}}
\vspace{-1mm}\caption{Plot of Model accuracy vs. number of features}
\label{features}
\end{figure}

\subsection{\textbf{Fault Detection (FD)}}

The FD utilizes the ICT network, which takes images of CQ features selected through the ReliefF algorithm as input. Fig.\ref{features} shows the relationship between the number of features and model accuracy across the three different matrix representations: RP, GAF, and MTF. As features are incrementally added through forward selection, accuracy improves for all models.  {The RP-based model reaches optimal accuracy with around 15 features, maintaining 100\% accuracy beyond this point, which supports the choice of using recurrence matrix with 15 features (5 per phase); adding further features offers no gain but increases computational time.}  {Table \ref{method} presents the $\mathcal{A}$ achieved with the three different imaging techniques when used as inputs to the ICT classifier. It can be observed that the RMCQ delivers the highest performance across all tasks, achieving 100\% $\mathcal{A}$ in fault detection, 90.23\% in fault location, and 98.61\% in phase selection. In comparison, GAF and MTF yield noticeably lower accuracies, particularly for fault location, underscoring the superiority of the proposed RMCQ representation.}
 Results indicate that applying SMOTE had minimal impact on accuracy, suggesting the method's robustness to class imbalance. The ICT network with RMCQ inputs achieves an $\mathcal{A}$ of 100\% both with SMOTE (balanced data) and without SMOTE (unbalanced data), underscoring it’s effectiveness in detecting faults across balanced and imbalanced datasets.

\begin{figure}[htpb]\vspace{-1mm}
\captionsetup{textfont=small}
\centerline{\includegraphics[width=3.85 in, height= 3.8 in]{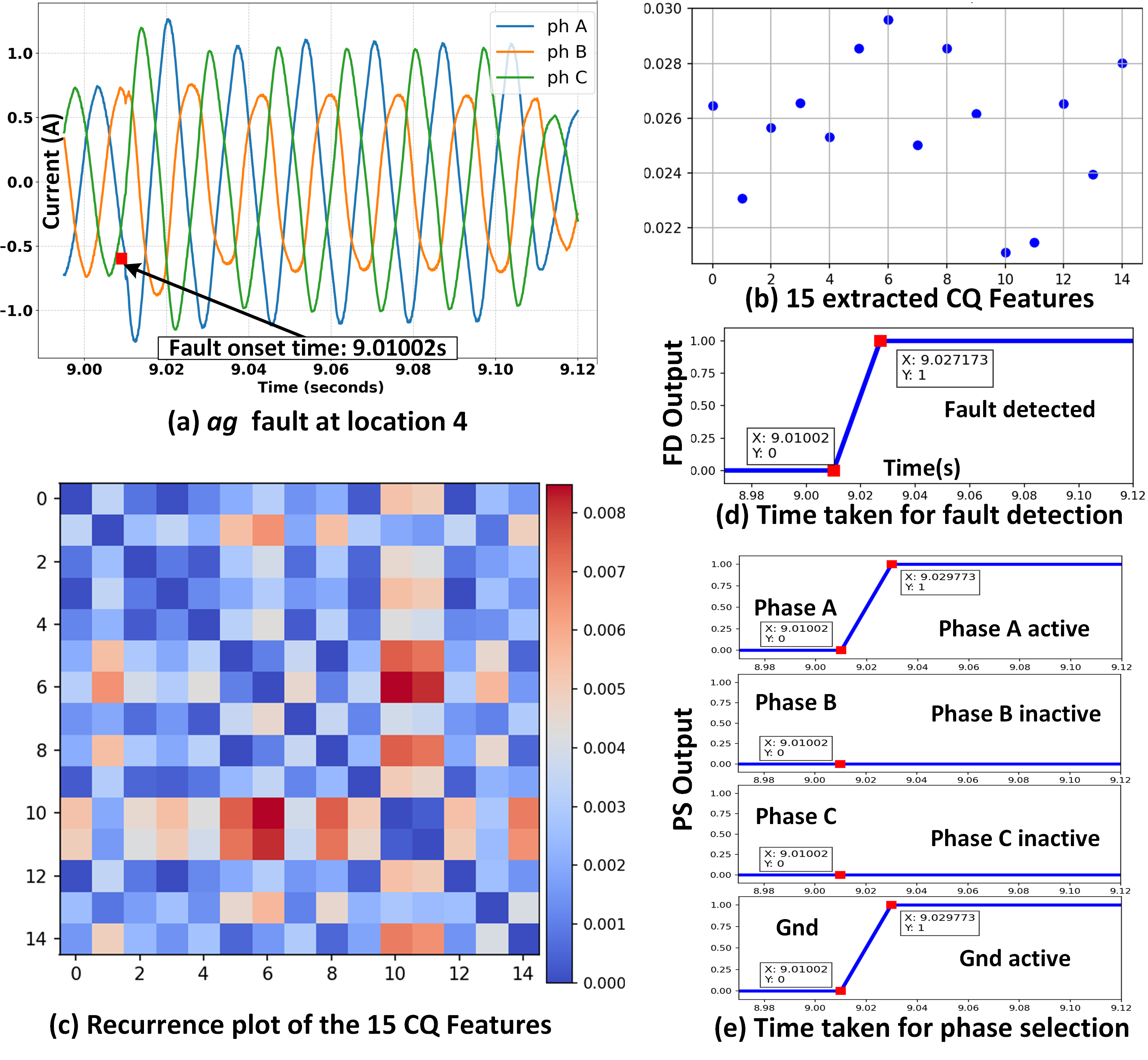}}
\caption{Plot showing (a) $lg$ fault ($ag$), (b) feature values, (c) recurrence matrix, (d) time taken by ICT network to detect fault with RMCQ as input,  (e) phase selection }
\label{lgfault}
\end{figure}

\begin{figure}[htpb]
\captionsetup{textfont=small}
\centerline{\includegraphics[width=3.85 in, height= 3.8 in]{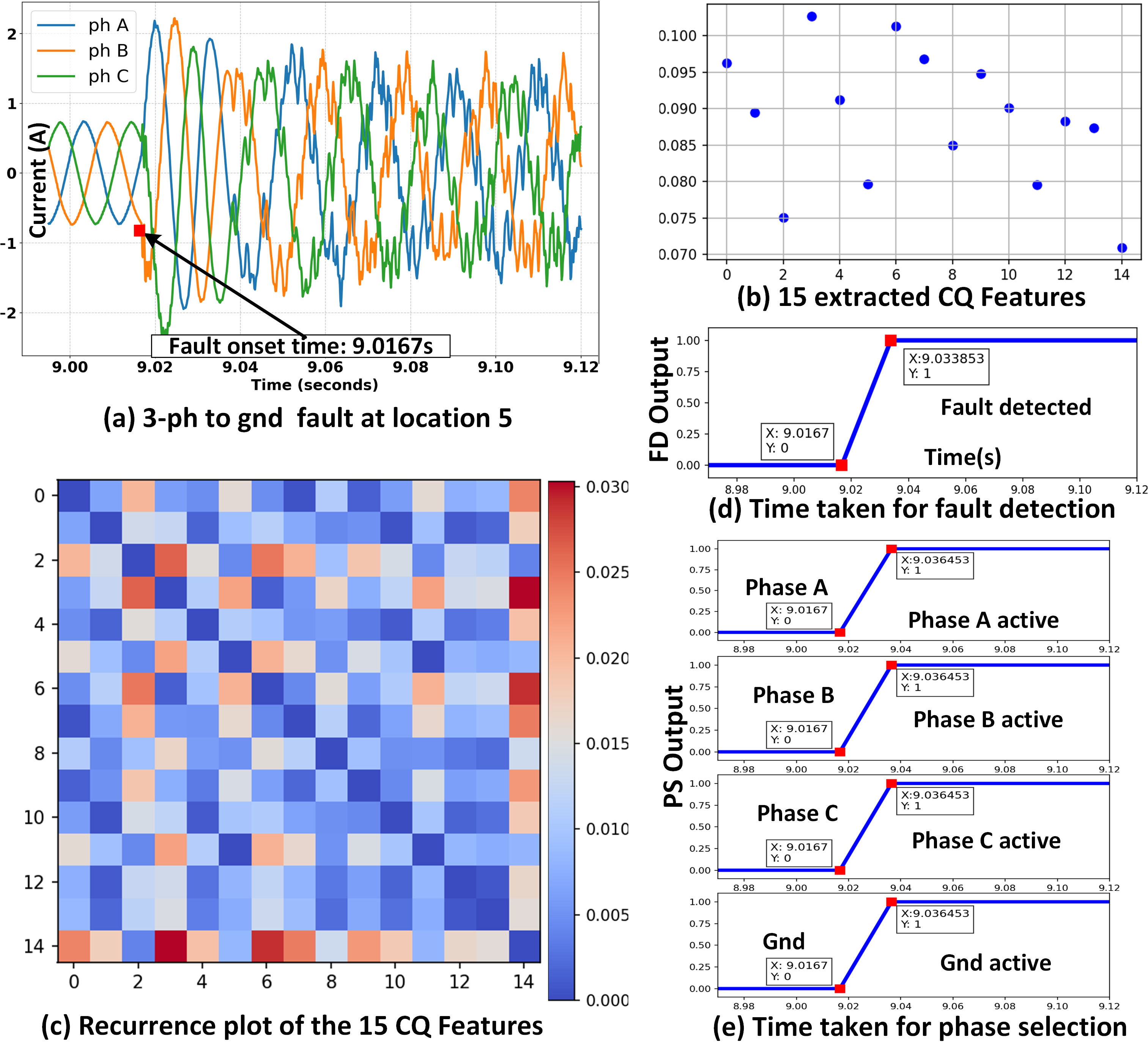}}
\caption{Plot showing (a) 3-phase to gnd fault ($abcg$), (b) feature values, (c) recurrence matrix, (d) time taken by ICT network to detect fault with RMCQ as input,  (e) phase selection }
\label{3phfault}
\end{figure}

\subsubsection{Performance with Internal Faults}
 {Figures \ref{lgfault} and \ref{3phfault}} illustrate the operation of the FD and PS scheme for unbalanced and balanced internal faults, respectively.  {Fig.\ref{lgfault}} shows a line-to-ground ($lg$) with $R_f$=0.01 $\Omega$ fault at point p4, the 15 CQ features extracted from 1 cycle of 3-phase currents, the 15×15 recurrence feature matrix of these CQs, the detection time taken by the ICT network with RMCQ as input, and the time required to identify faulty phases. For the $ag$ fault at 9.01002s, the FD changed status at 9.027173s and the PS status changes at 9.029773s. Similarly,  {Fig.\ref{3phfault}} depicts a 3-phase-to-gnd fault at point p5 with $R_f$=1 $\Omega$, along with the corresponding CQ features, recurrence matrix, fault detection time, and phase identification time. For the $abcg$ fault at 9.0167s, the FD changed status at 9.033853s and the PS status changes at 9.036453s. The recurrence matrix for  faults are larger values distributed throughout the matrix.

\begin{figure}[htpb]
\captionsetup{textfont=small}
\centerline{\includegraphics[width=3.85 in, height= 3.15 in]{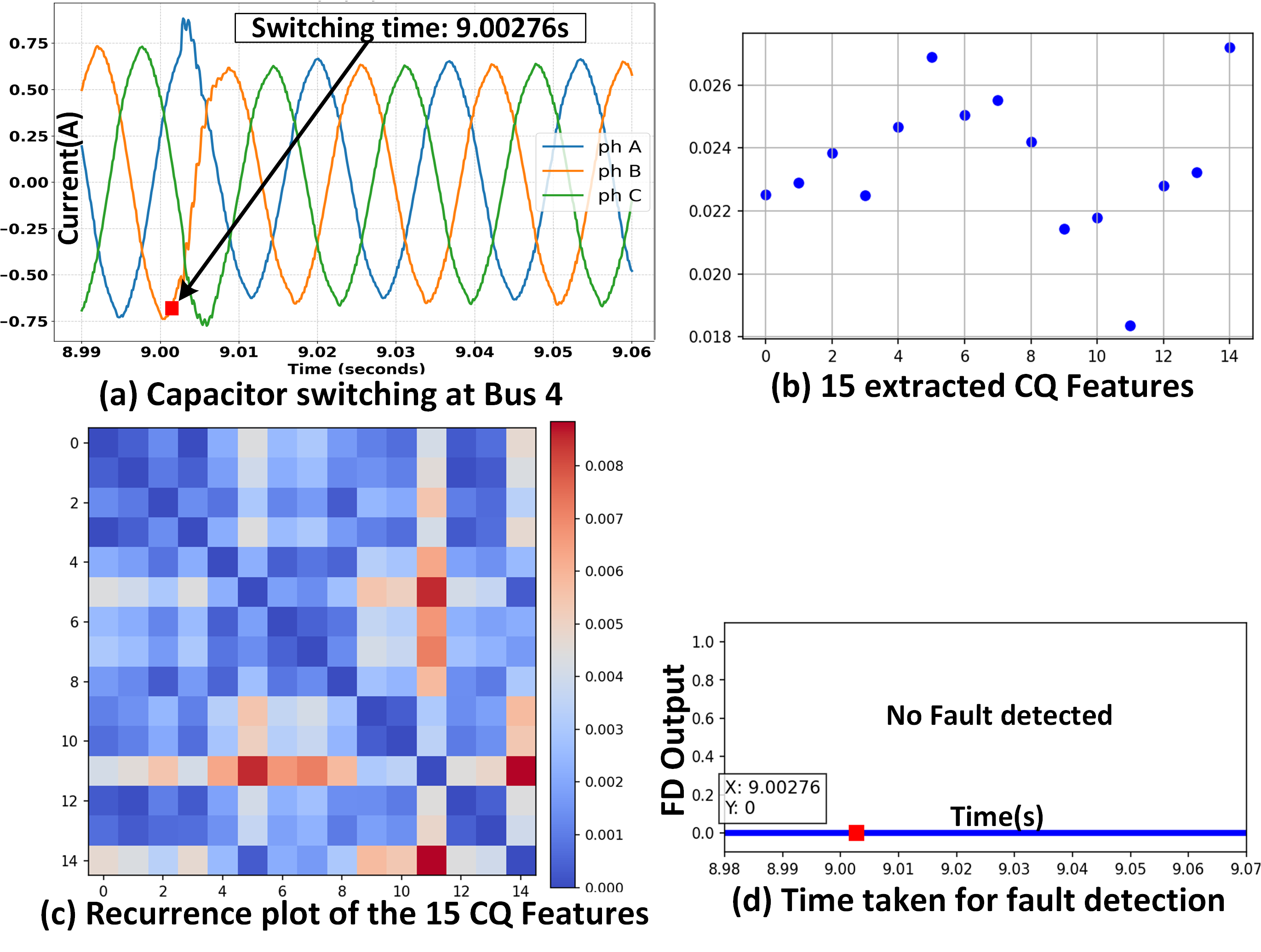}}
\caption{Plot showing (a) capacitor energization, (b) feature values, (c) recurrence matrix, and  (d) FD output}
\label{cap}
\vspace{-1mm}\end{figure}

\begin{figure}[htb]
\captionsetup{textfont=small}
\centerline{\includegraphics[width=3.85 in, height= 3.15 in]{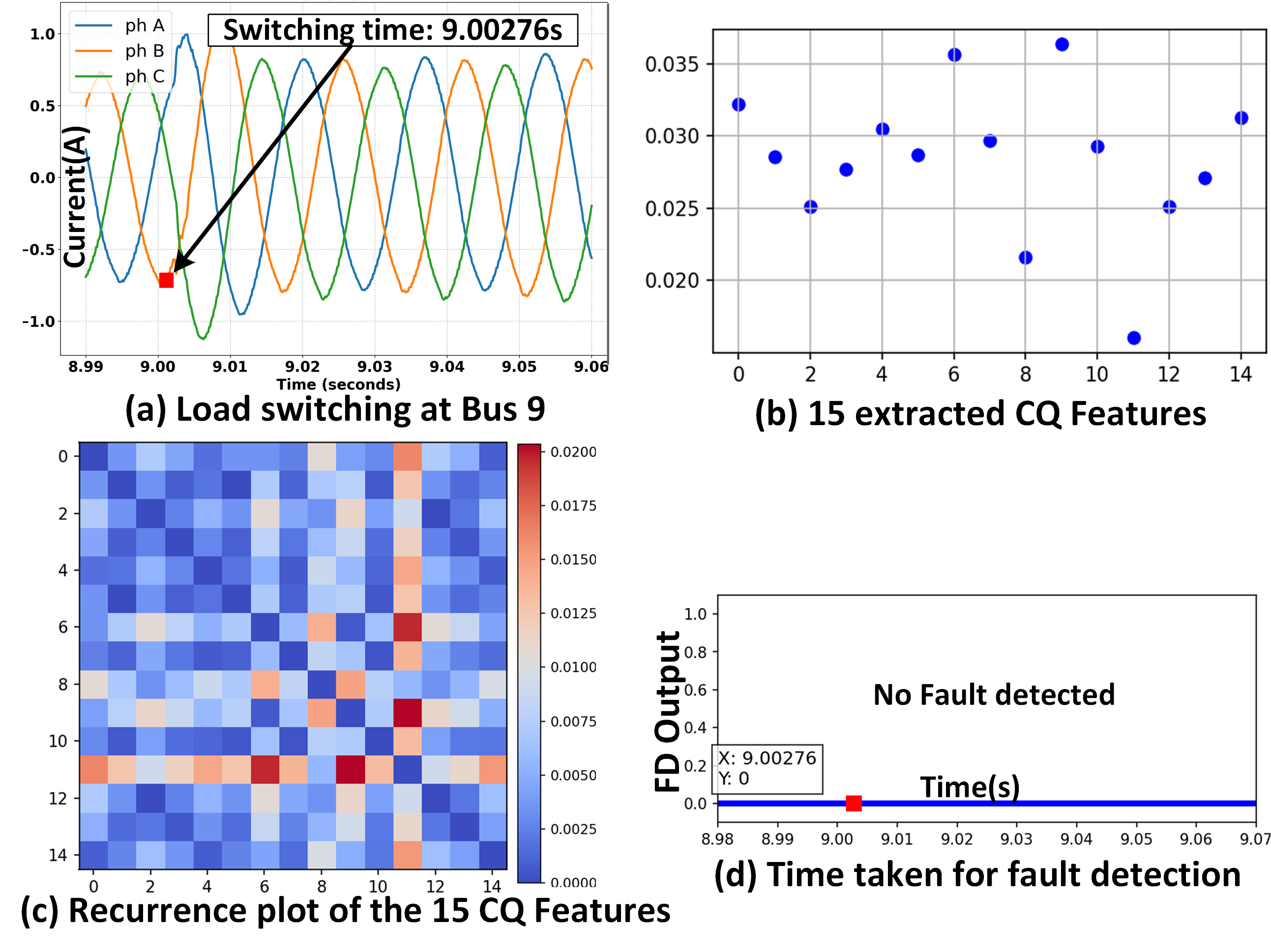}}
\caption{Plot showing (a) load switch, (b) feature values, (c) recurrence matrix, and  (d) FD output}
\label{load}
\vspace{-1mm}\end{figure}

\subsubsection{Performance with Other Transients}
  {Figures \ref{cap} and \ref{load}} present the FD scheme's response to capacitor and load switching events. They display the 3-phase currents, the 15 CQ features, the 15×15 recurrence feature matrix, and the ICT network response with RMCQ as input during capacitor and load switching, respectively. 500 MW, 200 MVAR load is switched onto bus-9 with the bus-3 generator engaged and priority mode P at 9.00276s. Again, 100 MVAR capacitor is switched onto bus-4 with the bus-3 generator engaged and priority mode P at 9.00276s. The recurrence matrix for other transients are clustered near zero and involves symmetric, low-magnitude values.

\definecolor{Silver}{rgb}{0.9,0.9,0.9}
\begin{table}[H]
\centering
\renewcommand{\arraystretch}{1.15}
\captionsetup{font=small} 
\captionof{table}{Performance of the three Time Series Imaging Technique}
\begin{tblr}{
rowsep=0.05 mm, 
  row{1} = {Silver},
  cell{1}{1} = {r=2}{},
  cell{1}{2} = {c=3}{},
  vlines,
  hline{1,3-6} = {-}{},
  hline{2} = {2-4}{},
}
Accuracy(\%)    & Imaging Technique &       &      \\
                & GAF               & MTF   & RP   \\
Fault Detection (FD)& 99.56             & 94.19 & 100.0  \\
Fault Location (FL)  & 83.91             & 54.51 & 90.23 \\
Phase Selection (PS) & 97.22                & 71.30    & 98.61 
\label{method}
\end{tblr}
\end{table}

%


\subsection{\textbf{Fault Location (FL)}}

Upon identifying a transient as a fault, the proposed scheme accurately locates the fault region. Fault simulations were conducted at eight positions: internal faults at positions p4 and p5 and external faults at positions p1, p2, p3, p6, p7, and p8. The ICT network, with RMCQ inputs, locates faults with an $\mathcal{A}$ of 90.23\% across 2880 cases (Table \ref{method}).  The confusion matrix is shown in Fig. \ref{cm}a.

\subsection{\textbf{Phase Selection (PS)}}

\begin{figure}[ht]
\captionsetup{justification=centering,textfont=footnotesize}
\centerline{\includegraphics[width=4.3 in, height= 2.2 in]{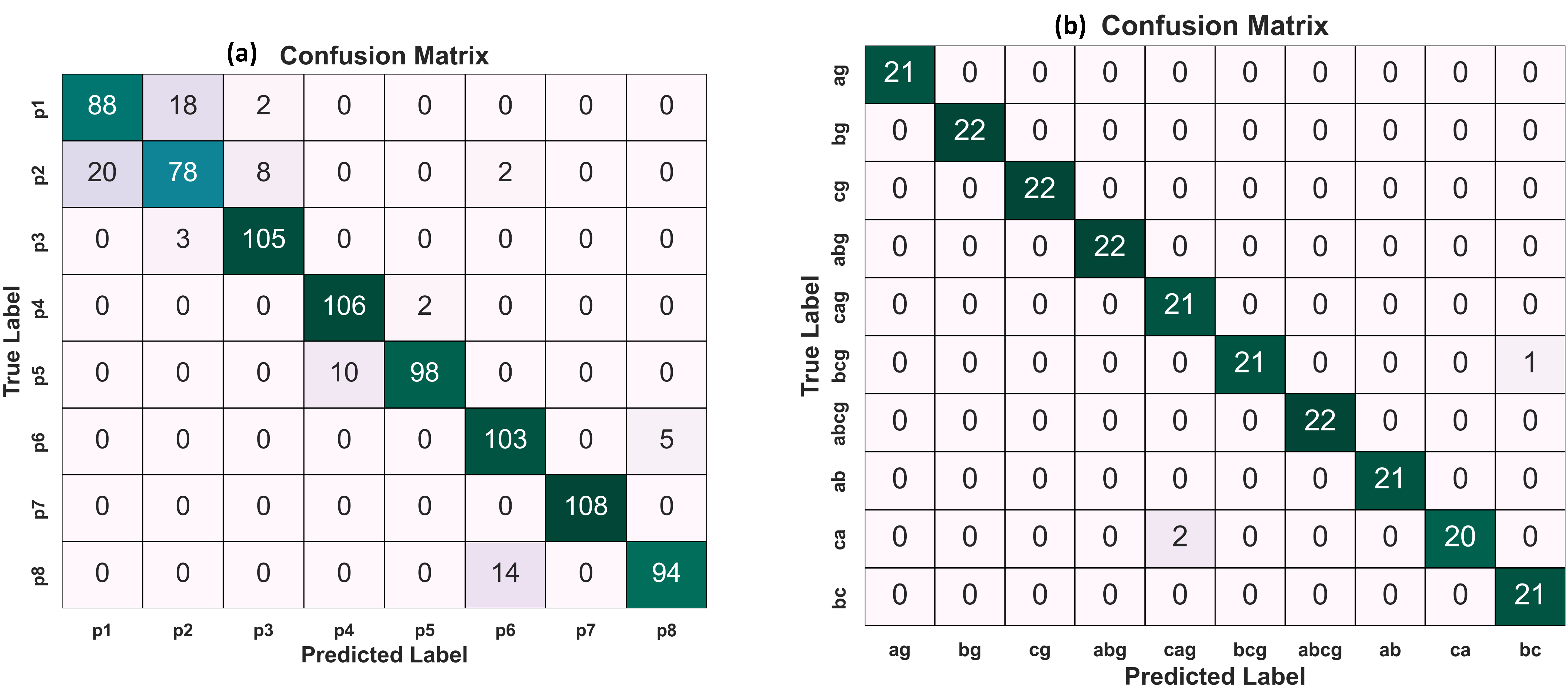}}
\caption{Confusion matrix for (a) Fault Location, (b) Phase selection}
\label{cm}
\vspace{-1mm}\end{figure}

Following FD and FL, the scheme proceeds to identify the faulty phase(s). The algorithm determines the involvement of phase a, phase b, phase c, or combinations such as phases ab, bc, ca, or abc. The ICT network achieves an $\mathcal{A}$ of 98.61\% with RMCQs as input in correctly identifying the faulty phase(s) on 720 fault cases at positions p4 and p5 (Table \ref{method}). The confusion matrix in Fig. \ref{cm}b illustrates the high accuracy across all phase types, indicating minimal misclassification between single-phase and multi-phase faults. This capability is crucial for effectively coordinating neighboring relays and reducing the risk of unintended outages. Figures \ref{lgfault}(e) and \ref{3phfault}(e) show the PS status and time taken for $ag$ and $abcg$ faults, respectively.

\section{Discussions on Validity of Proposed Scheme}
This section examines potential scenarios that may pose challenges to the proposed RMCQ-based fault detection (FD) scheme. The ability of the ICT to detect faults and switching transients is evaluated under these specific conditions. To facilitate this, the ICT network is trained and tested on 15×15 RMCQs on the 9-bus test system.

\subsection{\textbf{Effectiveness with Double Circuit Line}}
Ground distance relays face reliability challenges due to complexities introduced by the mutual coupling inherent in double-circuit TLs, which often necessitates advanced countermeasures \cite{double}. For this validation, a 100 km double-circuit TL, operating at 230 kV and 60 Hz, is modeled between buses $3_{SF}$ and 9 on the WSCC 9-bus system. The proposed FD scheme demonstrated robust performance by accurately distinguishing 1,260 fault events from 2,400 switching transients with $\mathcal{A}$ of 99.82\%. These faults were introduced at 50km from bus-$3_{SF}$ in P and Q modes, accounting for a range of fault impedances (0.01, 1, and 10 ohms), fault types (10 distinct configurations), and fault onset moments (21 intervals). 

\subsection{\textbf{Adaptability to  different power generation scenarios  }}
The performance of the proposed FD scheme was evaluated by altering the capacity of the PV system, adjusting the number of PV units from 400 to 300 and 500. This change in farm capacity introduces variations in power output, affecting the dynamics and fault characteristics within the system. The method demonstrated robust recognition capabilities, achieving  $\mathcal{A}$ of 99.8\% for 300 units and 99.5\% for 500 units. Validation was conducted using a dataset of 2,400 no-fault transients and 2,160 fault instances, where faults were simulated across different operating modes (P and Q), fault impedances (0.01, 1, and 10 ohms), and fault types (10 configurations). Six fault onset timings and positions p2, p5, and p6 were selected for fault introduction.

\subsection{\textbf{Effectiveness in presence of series compensation}}
 Thyristor Controlled Series Capacitors (TCSCs) used for power flow control, dynamic and transient stability, voltage stability, and damping oscillations can affect the operation of distance relays in the presence of IBRs \cite{sauviktcsc}. In the 9-bus test system, a TCSC (comprising a capacitor, inductor, and metal oxide varistor) is installed between bus-9 and bus-4 to provide up to 50\% compensation. By varying priority modes (P and Q), fault positions (p2, p5, p6), fault impedances (0.01, 1, 10 ohms), fault types (10 configurations), and 6 fault onset moments, 1080 faults and 2400 switching transients were simulated. The proposed method achieved an $\mathcal{A}$ of 99.5\% in identifying the transients and faults.

\subsection{\textbf{Resilience under Noisy Conditions}}
Measurement noise is a challenge for protection schemes in power systems, as it can lead to false tripping, missed fault detection, and coordination problems among protective devices. Sources of electromagnetic interference, such as nearby equipment and parallel power lines, contribute to noise in current waveforms. To evaluate the anti-noise capability of the proposed FD scheme, RMCQs extracted from 3-phase currents with SNRs  of 20, 30, and 40 dB were applied. Gaussian white noise was introduced into the 3-phase current signals, simulating realistic noise conditions encountered in field environments.  {The scheme’s $\mathcal{A}$ decreased from 100.0\% under noise-free conditions (SNR = $\infty$) to 84.0\% at an SNR of 20 dB, as shown in Table \ref{noise}. }
 This performance degradation under increasing noise highlights the importance of robust noise-resilient design in protection schemes.

\renewcommand{\arraystretch}{1.15}
\setlength{\tabcolsep}{6pt}
\begin{table}[H]
\centering
\footnotesize
\begin{minipage}{2.1in}

\captionof{table}{Effect of noise in phase current.}\label{noise}
\centering
\begin{tabular}{|l|c|}
\hline
 \rowcolor[rgb]{0.91,0.91,0.91}  {\textbf{SNR (dB)}} & $\mathcal{A(\%)}$ \\ \hline
$\infty$        &  100.0      \\ \hline
40        &  99.2         \\ \hline
30        &  93.4         \\ \hline
20        &  84.0        \\ \hline
\end{tabular}
\end{minipage}
\hspace{15mm}
\begin{minipage}{2.1in}
\centering

\captionof{table}{Effect of sampling rate of phase current.}\label{fs}

\begin{tabular}{|l|c|}
\hline
 \rowcolor[rgb]{0.91,0.91,0.91} \textbf{Sampling(kHz)} & $\mathcal{A(\%)}$ \\ \hline
7.68         &  100.0\%     \\ \hline
5.76        &  99.8\%       \\ \hline
5.12        &  99.8\%       \\ \hline
3.84        &  100.0\%      \\ \hline
\end{tabular}
\end{minipage}
\vspace{-1mm}
\end{table}

\subsection{\textbf{Resilience to various Sampling Rate \& Window Sizes}}
The relay’s performance in terms of speed and reliability is directly influenced by both the sampling frequency and the data window length. 
High sampling frequencies provide detailed insights into system behavior, capturing fast transients and high-frequency disturbances that are crucial for precise FD. However, lower sampling rates may fail to capture these nuances, potentially causing maloperation. Additionally, high sampling rates generate substantial data volumes, posing challenges for real-time processing, storage, and transmission. Therefore, balancing sampling frequency with data management capabilities is essential to optimize both performance and efficiency.
To examine the effect of sampling frequency, the proposed scheme was tested by sampling the 3-phase relay currents at various rates. Results indicate the efficacy of the proposed methods across different frequencies (Table \ref{fs}).

The choice of data window size is equally important, as it defines the time span of analyzed samples, affecting both temporal resolution and computational load. Window sizes of 0.5-cycle, 1-cycle, and 2-cycle were evaluated, and the proposed scheme maintained  $\mathcal{A}$ at 100\% across all these window sizes. 

 {The optimal parameter combination was found to be  3.84 kHz sampling frequency with  0.5-cycle window, which achieved 100\% $\mathcal{A}$ while ensuring fast detection suitable for protection applications.}

\subsection{\textbf{Effectiveness with High Impedance Faults}}

High impedance faults pose challenges for conventional protection systems due to their asymmetrical nature and low fault current levels, which can fall below the sensitivity range of distance or overcurrent relays \cite{hif}. To effectively model HIFs, a configuration consisting of two anti-parallel DC sources, controlled by two diodes and two variable resistors is used. In this model, the dynamic arc characteristics are captured through the variable resistors, the diodes control the current direction, and asymmetry in fault currents is represented by variations in the DC sources.
For this analysis, 370 HIFs were simulated at fault positions p4, p5, p6, p7, and p8 with faults applied in both P and Q modes and 37 distinct fault onset moments. Fault conditions were created using a phase-a to gnd fault with  impedances randomly set between 50 and 300 ohms, updated at 2 ms intervals. The FD scheme successfully differentiated HIFs from other switching transients with an $\mathcal{A}$ of 100\%.

\subsection{\textbf{Effectiveness with Cross-country and Evolving Faults}}

\begin{figure}[ht]\vspace{-1mm}
\vspace{-1mm}
\centering
\captionsetup{textfont=small}
\includegraphics[width=3.75in, height=2.05 in]{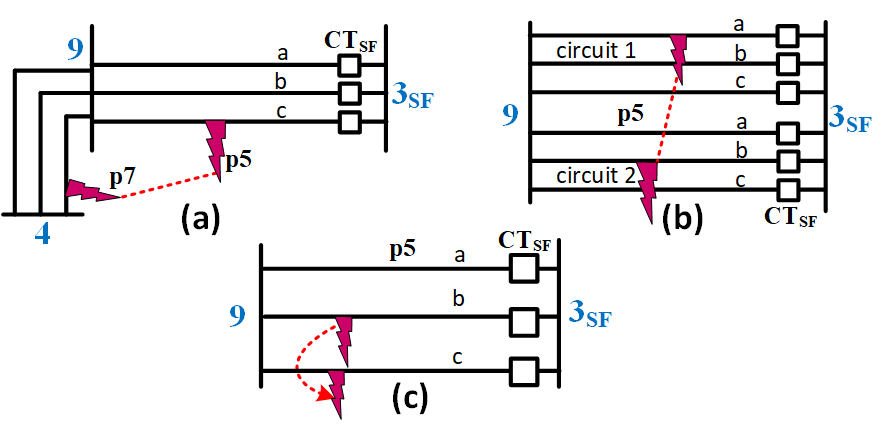}
\vspace{-1mm}
\caption{(a) Cross-country: $cg$ at p7 \& $cg$ at p5 at 9.001670s, (b) Cross-country: $ag$ (circuit 1) \& $bg$ (circuit 2) at p5 at 9.00167s, (c) evolving: $bg$ at 9.00167s converted to $bcg$ at 9.00334s at p5.}
\label{faults}
\vspace{-1mm}\end{figure}

The performance of distance relaying schemes can be significantly impacted by complex fault scenarios such as cross-country and evolving faults \cite{SWETAPADMA20151}. Cross-country faults involve simultaneous faults occurring at two different positions, either with the same or differing fault onset timings as shown in Fig. \ref{faults}(a,b) while evolving faults consist of primary and secondary faults occurring at the same position but with distinct onset timings (Fig. \ref{faults}(c)).

To evaluate the proposed scheme under cross-country fault conditions, 1,188 cases were simulated by introducing faults in both P and Q modes and varying fault onset moments across 11 intervals, with fault impedances of 0.01, 1, and 10 ohms. Two configurations were tested: (1) simultaneous $lg$ faults at separate positions, such as a $cg$ fault at position p7 and an $ag$, $bg$, or $cg$ fault at position p5 (see Fig. \ref{faults}(a)), and (2) simultaneous $lg$ faults in both circuit 1 and circuit 2 at the same position p5, such as an $ag$ fault in circuit 1 and an $ag$, $bg$, or $cg$ fault in circuit 2 (Fig. \ref{faults}(b)). The proposed RMCQ-based scheme accurately identified all cross-country fault cases, achieving an $\mathcal{A}$ of 100\%.

For evolving faults, 396 cases were simulated by introducing primary and secondary faults at the same position, p5, with varied onset moments, impedances, and priority modes (P and Q). In this scenario, $lg$ faults in one phase evolved into $llg$ faults involving two phases, such as an $ag$ fault evolving to $abg$ or $acg$, a $bg$ fault evolving to $abg$ or $bcg$, and a $cg$ fault evolving to $bcg$ or $acg$ (see Fig. \ref{faults}(c)). The scheme maintained an $\mathcal{A}$ of 100\% in detecting these evolving fault conditions.

\subsection{\textbf{Effectiveness with Remote faults}}
Traditional relays often encounter challenges with local/near-end and remote-end faults. Local faults can cause relay malfunctions due to high current magnitudes and low voltages, leading to CT saturation, which impairs FD. Conversely, remote faults present a different challenge: they may go undetected as both the voltage and current magnitudes can remain within nominal ranges.
In this study, 720 fault cases were simulated by altering parameters such as fault onset moments, fault impedances, modes (P and Q), and fault types. Positions p5 and p3 were chosen as the remote and local fault locations, respectively, to evaluate the scheme's adaptability across varying distances from the relay point. The proposed FD scheme  achieved an $\mathcal{A}$ of 100\% in correctly identifying both local and remote faults. 

\subsection{\textbf{Capability to handle CT Saturation Faults}}
During severe fault conditions, CT cores can become saturated, leading to distorted current signals that may negatively affect the performance of traditional protection algorithms \cite{ctsat}.
To evaluate the performance, the secondary burden of the CTs were increased to 20 ohms, inducing saturation under fault conditions. The scheme was then tested on a set of 1,760 fault cases across both P and Q modes, with fault impedances set to 0.01, 1, 2, and 10 ohms, covering 10 fault types and 11 different fault onset moments at fault positions p5 and p6 and 2,400 switching transient cases. The RMCQ-based FD scheme successfully identified all fault cases with an $\mathcal{A}$ of 100\%, demonstrating its robustness when CT saturation occurs.




\section{Fault Detection with Semi-Supervised Learning}\label{sec:SSL}
{	To mitigate the scarcity of labeled disturbance and fault data, a semi-supervised learning (SSL) framework was implemented using three paradigms: {label spreading (LS)}, {label propagation (LP)}, and {self-training (ST)}.  
	Each SSL variant acted as a \emph{teacher} that generated pseudo-labels for unlabeled samples, while the deep ICT network served as a \emph{student (head)} classifier trained on the fused labeled and pseudo-labeled dataset.  
	All experiments were conducted with unlabeled fractions $f_u \in \{0.2, 0.5, 0.8\}$ and $200$ training epochs for the ICT network to ensure convergence and stable temporal feature extraction.}

\subsection{\textbf{Label Spreading (LS)}}
{	LS constructs a similarity graph where edges encode feature-space proximity derived from recurrence-plot-transformed quantile features.  
	Labels are iteratively diffused through the graph as
	\begin{equation}
		F_{t+1} = \alpha S F_t + (1 - \alpha) Y,
	\end{equation}
	where $S$ is the normalized affinity matrix, $Y$ the initial label matrix, and $\alpha$ the clamping factor controlling the retention of original labels.  
	The parameter grid was
	$\alpha \in \{0.2, 0.6, 0.8\}, k \in \{5, 10, 20\},  \gamma \in \{0.001, 0.01, 0.1\}$
	corresponding to $k$-Nearest Neighbor (kNN) and Radial Basis Function (RBF) kernels.}

\subsection{\textbf{Label Propagation (LP)}}
{	LP uses the same graph but enforces {hard-clamping} on labeled nodes, retaining their initial labels during propagation:
	\begin{equation}
		F_{t+1} = S F_t, \quad F_i = Y_i,~\forall i \in \mathcal{L},
	\end{equation}
	where $\mathcal{L}$ denotes the labeled subset.  
	The same $k$ and $\gamma$ grids were employed.  
	While this constraint prevents label drift, it reduces adaptability under extreme scarcity.}

\begin{table*}[!t]
	\footnotesize
	\centering
	\caption{Performance of SSL paradigms across unlabeled fractions (ICT head).}
	\label{tab:ssl_results}
	\begin{tabular}{c c c c c c c c c c}
		\hline
	\rowcolor[rgb]{0.91,0.91,0.91}	$f_u$ & Method & Kernel & Parameters & F1 & AUPRC & AUROC & Recall & Precision & $\mathcal{A}$ \\
		\hline
		0.2 & LS & kNN & $k{=}10,\alpha{=}0.2$ & 0.992 & 0.999 & 0.999 & 0.985 & 0.999 & 0.992 \\
		0.2 & LP & kNN & $k{=}20$ & 0.985 & 0.999 & 0.999 & 0.982 & 0.989 & 0.987 \\
		0.2 & ST & -- & threshold=0.7 & 0.990 & 0.998 & 0.998 & 0.982 & 0.999 & 0.991 \\
		\hline
		0.5 & LS & kNN & $k{=}10,\alpha{=}0.2$ & 0.980 & 0.998 & 0.998 & 0.978 & 0.982 & 0.982 \\
		0.5 & LP & kNN & $k{=}10$ & 0.974 & 0.996 & 0.997 & 0.958 & 0.990 & 0.977 \\
		0.5 & ST & -- & threshold=0.7 & 0.970 & 0.991 & 0.989 & 0.959 & 0.980 & 0.973 \\
		\hline
		0.8 & LS & kNN & $k{=}10,\alpha{=}0.2$ & 0.960 & 0.985 & 0.989 & 0.971 & 0.950 & 0.963 \\
		0.8 & LP & kNN & $k{=}10$ & 0.949 & 0.983 & 0.986 & 0.979 & 0.920 & 0.952 \\
		0.8 & ST & -- & threshold=0.9 & 0.930 & 0.963 & 0.966 & 0.914 & 0.946 & 0.938 \\
		\hline
	\end{tabular}
\end{table*}

\subsection{\textbf{Self-Training (ST)}}
{	A RF base classifier ($400$ trees) was first trained on the labeled subset and then iteratively predicted pseudo-labels for unlabeled samples whose confidence exceeded a threshold $\tau$.  	These pseudo-labeled samples were appended to the training set until convergence.  The threshold grid was $\tau \in \{0.7, 0.8, 0.9\}$.	}

{	After each SSL stage (LS, LP, or ST), pseudo-labels were merged with true labels:
	\begin{equation}
		y_{\text{student}} =
		\begin{cases}
			y_{\text{pseudo}}, & y=-1,\\[3pt]
			y_{\text{true}}, & \text{otherwise},
		\end{cases}
	\end{equation}
	and the ICT network was trained on $(X, y_{\text{student}})$ with batch size $64$.} {	
	Each configuration was evaluated using precision ($P$), recall ($R$), F1-score, area under the ROC (AUROC), area under the Precision–Recall curve (AUPRC), and accuracy ($\mathcal{A}$). 
	Results are summarized in Table~\ref{tab:ssl_results}. Across all unlabeled fractions ($f_u=0.2-0.8$), the SSL framework maintained near-supervised
	performance (F1: 0.992$\rightarrow$0.960, $\mathcal{A}$: 0.992$\rightarrow$0.963).
	LS with ICT proved the most robust and data-efficient approach under severe label shortages.}

\begin{figure}[ht]
	\centering
	\captionsetup{justification=centering,textfont=small}
	\includegraphics[width=3.75 in, height=3.35 in]{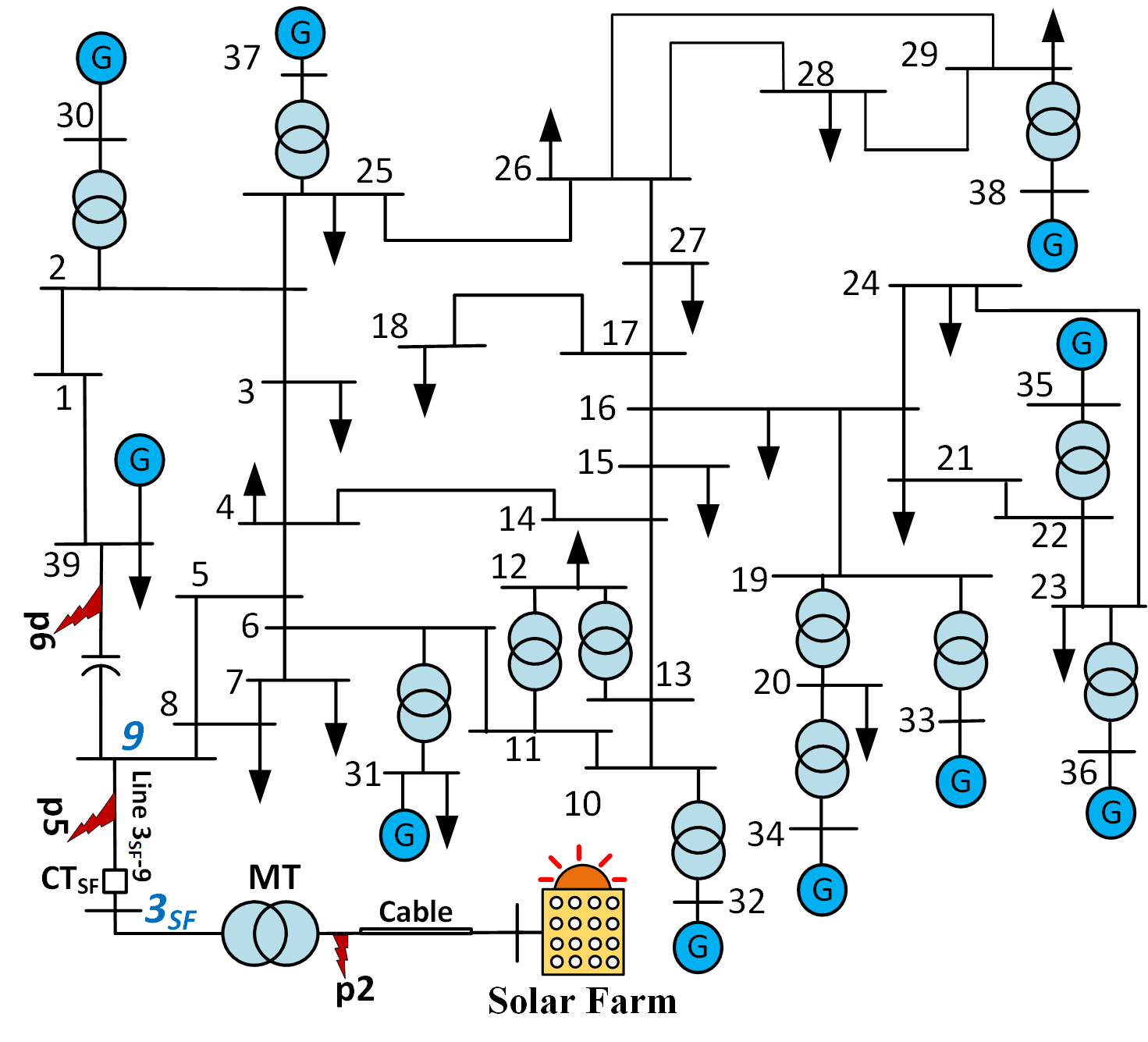}
	\vspace{-0.1cm}
	\caption{New England IEEE 39-Bus System with PV Plant at bus-9.}
	\label{39busPV}
	\vspace{-0.1cm}
\end{figure}

\section{Validation on IEEE 39-Bus Test System}
To assess the performance of the proposed protection scheme under realistic and complex power grids, capture inter-area power flows and dynamic interactions between regions, include wide variety of operating conditions, and to make it scalable and credible, the proposed method is evaluated on a widely used bigger test system.
The IEEE 39-bus system \cite{39bussystem}, commonly referred to as the 10-machine New England Power System, is used for validation (Fig. \ref{39busPV}). Table \ref{parameters1} with fault positions: p2, p5, and p6; and Table \ref{parameters2} outlines the parameters and their values used to simulate various scenarios. The RMCQ-based FD method was applied to the 39-bus system, achieving an $\mathcal{A}$ of 99.6\% across 1,080 fault cases and 2,400 non-fault cases. 
\vspace{-1mm}

\begin{figure}[ht]
\centering
\captionsetup{justification=centering,textfont=small}
\includegraphics[width=4.9 in, height=2.0 in]{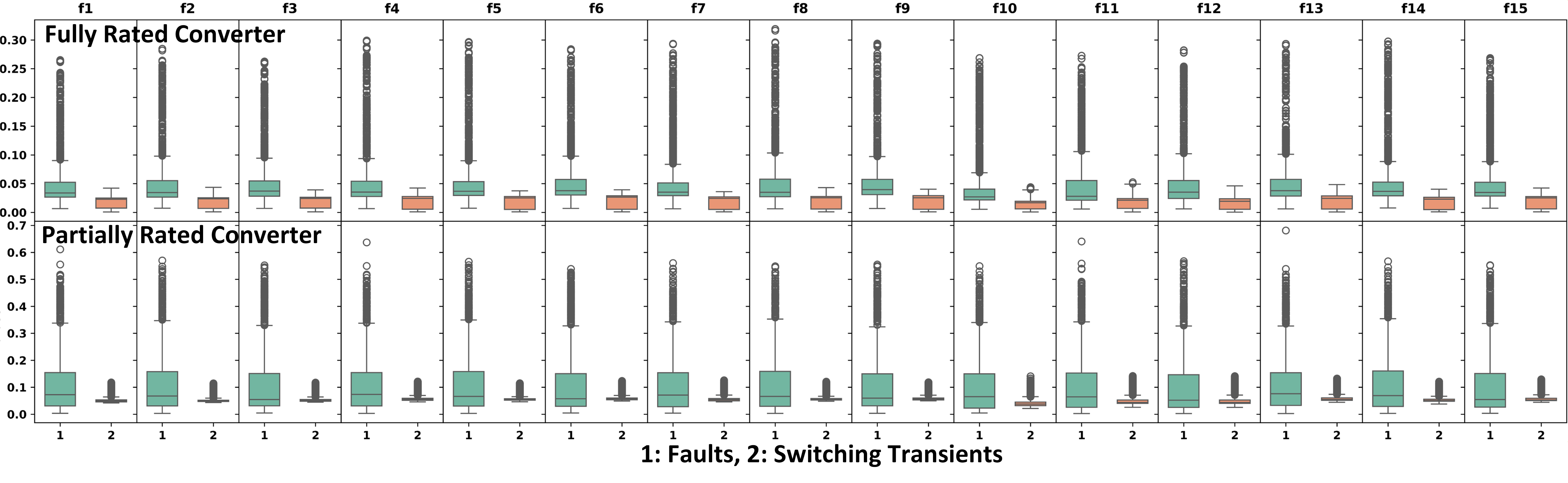}
\vspace{-0.1cm}
\caption{Boxplots for the 15(5$\times$3) CQ features for fully rated and partially rated converters.}
\label{wfpv}
\vspace{-1mm}\end{figure}

\begin{table}[ht]
\centering
\renewcommand{\arraystretch}{1.1}
\setlength{\tabcolsep}{3 pt}
\caption{A Comparative Review of Recent Articles}\label{comparison}
\begin{tabular}{|>{\itshape}l|c|c|c|c|c|} \hline
\rowcolor[rgb]{0.91,0.91,0.91} Reference                 & \begin{tabular}[c]{@{}l@{}}\cite{singh}\\2024 \end{tabular}  &\begin{tabular}[c]{@{}l@{}}\cite{AKTER2022}\\2022 \end{tabular}  &  \begin{tabular}[c]{@{}l@{}}\cite{GHORBANI2023}\\2023 \end{tabular}& \begin{tabular}[c]{@{}l@{}}\cite{pallavpv}\\2024 \end{tabular} & \begin{tabular}[c]{@{}l@{}}Current \\ approach\end{tabular} \\ \hline
Method                                                                      & \begin{tabular}[c]{@{}l@{}}current\\mag. ratio \end{tabular}& impedance & \begin{tabular}[c]{@{}l@{}}+tive seq.\\network\end{tabular} & \begin{tabular}[c]{@{}l@{}}CLT, \\RF \end{tabular}& \begin{tabular}[c]{@{}l@{}}RMCQ,\\ICT \end{tabular}                                             \\ \hline
Signals                                                               &  $I$& $ V$ \& $I$ & $ V$ \& $I$ & $I$  &  $I$                                                     \\ \hline
\begin{tabular}[c]{@{}l@{}}Single/double end\end{tabular}                & double & single &single &  single& single                                                     \\ \hline
\begin{tabular}[c]{@{}l@{}}System f (Hz)\\ Samp. f (kHz)\end{tabular}& 50,1 & 60,1 &  60,1.2&  60,7.68& 60,3.84                                                        \\ \hline
Time delay(ms)                                                              & 10 & 16.67 &  -&  18.5 &  12.82                                                      \\ \hline
\begin{tabular}[c]{@{}l@{}}FD $\mathcal{A}$ (\%)\end{tabular}& 100 & 100 &  100&  98.2& 100                                                       \\\hline
\rowcolor[rgb]{0.91,0.91,0.91} 
\multicolumn{6}{|c|}{Impact of Scenarios Considered} \\ \hline
High impedance faults                                                              & \checkmark & - & \checkmark & \checkmark & \checkmark                                                        \\ \hline
Noise                                                             & \checkmark & \checkmark & \checkmark & \checkmark & \checkmark                                                        \\ \hline
\begin{tabular}[c]{@{}l@{}} Double ckt. lines\end{tabular}       & - & - & - & \checkmark  & \checkmark                                                        \\\hline
\begin{tabular}[c]{@{}l@{}}Farm capacity\end{tabular}    &- & - & - & \checkmark  & \checkmark                                                        \\\hline
\begin{tabular}[c]{@{}l@{}}Cross country faults\end{tabular}    &-  & - & - & \checkmark  & \checkmark                                                        \\\hline
\begin{tabular}[c]{@{}l@{}}Series compensation\end{tabular}    &-  & - & - & \checkmark  & \checkmark                                                        \\\hline
\begin{tabular}[c]{@{}l@{}}Evolving faults\end{tabular}         & - & \checkmark & \checkmark & \checkmark  & \checkmark                                                        \\\hline
\begin{tabular}[c]{@{}l@{}} Sampling freq.\end{tabular}         & - & - & - & \checkmark  & \checkmark                                                        \\\hline
\begin{tabular}[c]{@{}l@{}}Data size\end{tabular}             & - & - & - & \checkmark & \checkmark                                                        \\\hline
\begin{tabular}[c]{@{}l@{}}CT saturation\end{tabular}           & - & \checkmark & \checkmark & \checkmark & \checkmark                                                        \\\hline
\begin{tabular}[c]{@{}l@{}}Load addition\end{tabular}          & \checkmark & \checkmark & - & \checkmark & \checkmark                                                        \\\hline
\begin{tabular}[c]{@{}l@{}}Capacitor energization\end{tabular}     & \checkmark & - & - & \checkmark & \checkmark                                                       \\ \hline
\begin{tabular}[c]{@{}l@{}}Remote faults\end{tabular}     & - &  \checkmark & - & \checkmark & \checkmark                                                       \\ \hline
\end{tabular}
\end{table}

\section{Validation with Partially Rated Converters}
The proposed protection scheme is further validated in the presence of a partially rated converter by integrating a type-3 wind farm. Type-3 wind farms exhibit fault characteristics, system dynamics, and control mechanisms distinct from fully rated converter systems \cite{nrel_report}, such as those found in PV farms. In this study, a 200MW type-3 wind farm replaces the PV farm in the IEEE 9-bus test system.
To evaluate the scheme, 2880 fault scenarios are simulated, considering wind speeds of 8m/s and 11m/s. The simulations account for variations in fault impedance (0.01, 1, and 10$\Omega$), fault types (10 distinct types), and fault onset moments (6 intervals in one cycle) at eight fault locations in the system. Additionally, 2400 switching scenarios are simulated, including capacitor and load switching events, using the parameters defined in Table \ref{parameters2} at wind speeds of 11m/s and 22m/s. The RMCQ-based FD scheme differentiated faults from the switching events with an $\mathcal{A}$ of 99.6\%, demonstrating its applicability to both fully rated and partially rated converters. Fig.\ref{wfpv} presents the boxplots of the 15 CQ features for faults and switching transients. The partially rated converter exhibits a wider interquartile range and overall range, indicating higher variability in feature values compared to the fully rated converter.


\section{Comparative Evaluation}
Table \ref{comparison} provides a summary of recent studies on the protection of transmission lines connected to PV systems, evaluating how these methods perform across different scenarios to determine the robustness of each approach. Although these methods show strong performance in specific scenarios, many critical cases remain unexplored.

The proposed method achieves a runtime similar to those reported in these studies.  {According to IEEE Std C37.243 \cite{standard2}, the end-to-end communication delay consists of relay interface delay, fiber-optic propagation delay, and multiplexer delays. The relay interface delay is typically 1–5 ms, the fiber propagation delay is 5 µs/km, and multiplexer delays are: 0.37 ms for a substation multiplexer, 1–100 ms for a telco channel bank, and 0.2 ms for SONET/SDH multiplexers. In a single-ended implementation, as in the proposed algorithm, these communication-related delays are absent.} The intelligent protection system, utilizing RMCQs, depends on data processing and inference times. Computing 15 RMCQs requires 4.0 ms, while the ICT ML model takes 0.48 ms to process new data. Consequently, the total runtime for 1/2 cycle of data is 12.82 ms (8.34 ms + 0.48 ms + 4.00 ms) and for 1 cycle is 21.15ms.

{\textit{InceptionTime  vs. Transformers:}
A comparative study is also conducted between the proposed ICT classifier and local-sparse transformer (LST) \cite{local} using identical CQ$\rightarrow$RP feature pipelines and train–test partitions. 
Both architectures achieved near-ceiling classification performance ($\mathcal{A}$ $\approx 100\%$); however, they exhibited distinct differences in computational latency.
McNemar’s test on the test set (30 \% of 5280 = 1584) showed no statistically significant difference (\textit{p}=1.0), indicating that neither model holds a decisive advantage in overall accuracy.
On the test set, ICT required approximately 2.0s of total inference time, compared to 6.5s for LST, primarily due to its convolutional structure and the absence of explicit attention computations.
Moreover, ICT trained $\sim\!25\times$ faster than LST and employed far fewer parameters, resulting in a smaller and more efficient model footprint..

\section{CONCLUSION}

 Maintaining dependability for internal faults and security against external faults and transients can be challenging for TLs linked to large-scale PV farms. 
The proposed RMCQ-based InceptionTime intelligent protection scheme proves dependable in detecting internal faults, cross-country faults, evolving faults, faults with CT saturation, and double circuit line faults, while remaining sensitive to low current conditions associated with high impedance faults in the IEEE 9-bus system. It ensures security during events such as capacitor bank energization and load addition. 
 {This method is robust against variations in PV capacity, data window size, sampling rates, measurement noise, and with series compensation. However, the method shows sensitivity to strong noise conditions (e.g., 20 dB SNR), and its effectiveness in real-world deployment may be affected by the scarcity of labeled fault data.}
It is also validated on the IEEE 39-bus system and with partially rated converters. The results indicate that the RMCQ-based system delivers comprehensive protection for TLs connected to bulk PV farms, ensuring dependability, security, robustness, and rapid response. {Moreover, the integration of semi-supervised learning strategies (label spreading, label propagation, and self-training) with the InceptionTime architecture effectively bridges the gap between supervised and unlabeled learning, ensuring high accuracy under label-scarce conditions.}
Further, the approach relies solely on locally measured data, eliminating the need for remote-end communication devices.

\bibliographystyle{elsarticle-num}
\small{
\bibliography{references}}

\end{document}